# TENET: A Framework for Modeling Tensor Dataflow Based on Relation-centric Notation

Liqiang Lu[1]
*Peking University*
liqianglu@pku.edu.cn

Naiqing Guan[1+]
*Peking University*
*University of Toronto*
naiqing.guan@mail.utoronto.ca

Yuyue Wang
*Peking University*
wangyuyue@pku.edu.cn

Liancheng Jia
*Peking University*
jlc@pku.edu.cn

Zizhang Luo
*Peking University*
semiwaker@pku.edu.cn

Jieming Yin
*Lehigh University*
yin@lehigh.edu

Jason Cong
*University of California at Los Angeles*
cong@cs.ucla.edu

Yun Liang*
*Peking University*
ericlyun@pku.edu.cn

***Abstract*—Accelerating tensor applications on spatial architectures provides high performance and energy-efficiency, but requires accurate performance models for evaluating various dataflow alternatives. Such modeling relies on the notation of tensor dataflow and the formulation of performance metrics. Recent proposed compute-centric and data-centric notations describe the dataflow using imperative directives. However, these two notations are less expressive and thus lead to limited optimization opportunities and inaccurate performance models.**

**In this paper, we propose a framework TENET that models hardware dataflow of tensor applications. We start by introducing a relation-centric notation, which formally describes the hardware dataflow for tensor computation. The relation-centric notation specifies the hardware dataflow, PE interconnection, and data assignment in a uniform manner using relations. The relation-centric notation is more expressive than the compute-centric and data-centric notations by using more sophisticated affine transformations. Another advantage of relation-centric notation is that it inherently supports accurate metrics estimation, including data reuse, bandwidth, latency, and energy. TENET computes each performance metric by counting the relations using integer set structures and operators. Overall, TENET achieves 37.4% and 51.4% latency reduction for CONV and GEMM kernels compared with the state-of-the-art data-centric notation by identifying more sophisticated hardware dataflows.**

## I. Introduction

Tensor operations have been increasingly deployed in many applications, such as data analysis, machine learning, and hydrodynamics simulation [1, 2, 16, 23, 34, 46, 47, 53]. In recent years, spatial architectures have emerged as a promising way to accelerate tensor operations due to their high performance and energy-efficiency [8, 10–12, 15, 18, 19, 21, 22, 26, 30, 36, 40, 42–44, 48, 57]. A typical spatial architecture is composed of a processing element (PE) array and a scratchpad memory. PEs are connected via an on-chip interconnect that enables efficient data reuse.

The main characteristic of spatial architectures is the diverse hardware dataflow alternatives. Tensor operations are usually described using a loop nest. Specific to tensor operation, a hardware dataflow describes 1) the assignment of loop instances to the PE array and 2) the execution sequence of these loop instances in the PEs. Hardware dataflow is critical for achieving high throughput and low latency, because it determines PE utilization, data access patterns, and on-chip bandwidth requirement. Different tensor computation prefers different hardware dataflows. For example, Google's Tensor Processing Unit (TPU) [22] connects PEs using systolic dataflow, where each PE is responsible for one multiply-and-accumulate operation. While Cambircon [31] connects PEs via a multicast communication network, in which each PE performs a dot-product. Other spatial architectures, like DySER [19] and Plasticine [42], integrate PEs and their interconnect in a flexible manner, and hence can support a wider range of applications.

Despite the fact that various dataflows have been practically implemented in modern tensor accelerators, a formal notation is still strongly desired to represent hardware dataflow. *Ideally, a notation should be able to cover the complete dataflow design space systematically, as well as facilitate simple and accurate performance modeling.* State-of-the-art techniques represent hardware dataflow using either compute-centric [39, 56] or data-centric notations [24, 25]. However, both notations have limitations. First, these notations are less expressive and they can only represent a subset space of hardware dataflows. Using these notations, architects are provided with an incomplete space and limited optimization opportunities. For example, both notations fail to describe hardware dataflows that require skewing of loop iterations and tensors. Such dataflows needs affine loop transformation [28, 29] to enable sophisticated mapping of loop instances to spatial architectures. Second, both notations fail to support accurate performance analysis. The compute-centric notation does not directly model data transfer and reuse, and thus lacks a detailed performance model. The data-centric notation is integrated with the MAESTRO model, which outputs the reuse, latency, and energy for a given dataflow [25]. However, MAESTRO models these metrics by calculating the polynomials of parameters, which might not be accurate for tensors whose dimensions cannot be explicitly specified by the data-centric primitives.

[1]These authors contributed equally.
+Work done while the author was a student at Peking University.
*Corresponding Author

TABLE I: Comparison between four notations

| Features | | Computation-centric Timeloop [39] | Computation-centric Interstellar [56] | Data-centric MAESTRO [24, 25] | STT [4, 9, 28, 54] | Relation-Centric TENET |
|---|---|---|---|---|---|---|
| Express dataflow | Instance execution sequence | loop order | loop order | temporal maps sequence of maps | time-stamp vector | multi-dim time-stamp |
| | PE workload assignment | `parallel` directive | `unroll` primitives | spatial maps | space-stamp matrix | multi-dim space-stamp |
| | Affine loop transformation | × | × | × | ✓ | ✓ |
| | Spatial architectures | ✓ | ✓ | ✓ | × | ✓ |
| Performance modeling | PE interconnection | × | × | × | × | ✓ |
| | Precise reuse analysis | × | × | × | × | ✓ |
| | Data assignment analysis | × | ✓ | ✓ | × | ✓ |
| | Bandwidth analysis | × | ✓ | ✓ | × | ✓ |
| | Latency / energy modeling | ✓ | × | ✓ | × | ✓ |
| | General tensor apps | × | × | × | ✓ | ✓ |

In this paper, we propose TENET, a framework that models the dataflow of tensor applications on spatial architectures. The key component of TENET is the *relation-centric notation*, which formally describes the hardware dataflow for tensor computation. Specifically, we formally define the relations between 1) the loop instances and the PEs that perform the computation, 2) the loop instances and their execution sequence in the PEs, 3) PEs and the corresponding assigned tensor elements, and 4) PEs that are connected with interconnection network. The first two relations determine where and when the loop instances are executed. The third relation models where and when a tensor element is accessed. The last relation describes how the tensor elements traverse across PEs, e.g., systolic array, reduction tree. By putting these relations together in a uniform manner, we can precisely model the tensor computation on spatial architectures, data assignment to PEs, and data movement between PEs.

The relation-centric notation allows the users to describe the complex hardware dataflows using relations only. When representing these relations, TENET supports all linear transformations that map loop instances to spatial architecture spatially and temporally. This leads to a complete design space of hardware dataflows. The relation-centric notation also inherently supports simple and accurate modeling of various important performance metrics. These metrics are crucial for evaluating different hardware dataflow design alternatives. All four types of relations above can be mathematically represented as a set of pairs. Based on such structural representation, performance metrics can be easily computed using integer set operators. Overall, TENET is able to estimate various hardware metrics, including data reuse, latency, PE communication bandwidth, and on-chip memory bandwidth.

The contributions of this work are as follows. First, we propose relation-centric notation for modeling the hardware dataflow of tensor computation. Relation-centric notation is more expressive than compute-centric and data-centric notations. By representing the dataflow, data assignment, and interconnection as relations uniformly, relation-centric notation forms the complete design space of hardware dataflows, which provides more optimization opportunities.

Second, we introduce performance models that can accurately calculate various hardware metrics. This is naturally supported by the structural representation of relation-centric notation. We first formulate three basic volumes that describe the overall data size, the reused data size, and the minimum data size that needs to be transferred between PEs and scratchpad. We then use these volumes to derive various hardware metrics.

Third, we systematically analyze and compare different notations in terms of expressiveness and performance modeling. Results show that TENET achieves 37.4% and 51.4% latency reduction for Conv and GEMM kernels by identifying more sophisticated dataflows compared to the state-of-the-art data-centric notation. The source code of TENET is publically available in Github (https://github.com/pku-liang/TENET).

## II. Background

### A. Spatial Architectures

Spatial architectures are a class of architectures that feature a set of processing element (PE), interconnection between PEs and memory hierarchy [7, 17, 19, 36, 40, 42, 48]. Each PE contains arithmetic logic units (ALUs) that can be configured by specific instructions. The PE also contains register files for data storage. The interconnection between PEs can effectively increase data reuse opportunities, which further reduces the bandwidth requirement. Generally, spatial architectures have three levels of memory hierarchy, i.e., PE register level, on-chip scratchpad, and off-chip memory. For simplicity, we make the following two assumptions when modeling hardware behaviors. First, the ALU has the ability to perform one multiply-and-accumulate (MAC) operation. Second, data transfer through the interconnect between adjacent PEs takes one cycle.

We use the term *hardware dataflow* to refer to the implementation of a specific tensor application on a spatial architecture. Tensor applications are usually described using loop nests. Specific to tensor applications, the dataflow is represented from two aspects: 1) the PE where a loop instance is executed, 2) the execution sequence of loop instances in the PEs. When designing the dataflow, data reuse is a critical factor in order to achieve high performance and low energy, which can be further categorized into temporal reuse and spatial reuse [10, 11, 24, 56]. Temporal reuse happens when the same data is reused at different cycles, while spatial reuse happens when the same data is reused at different PEs.

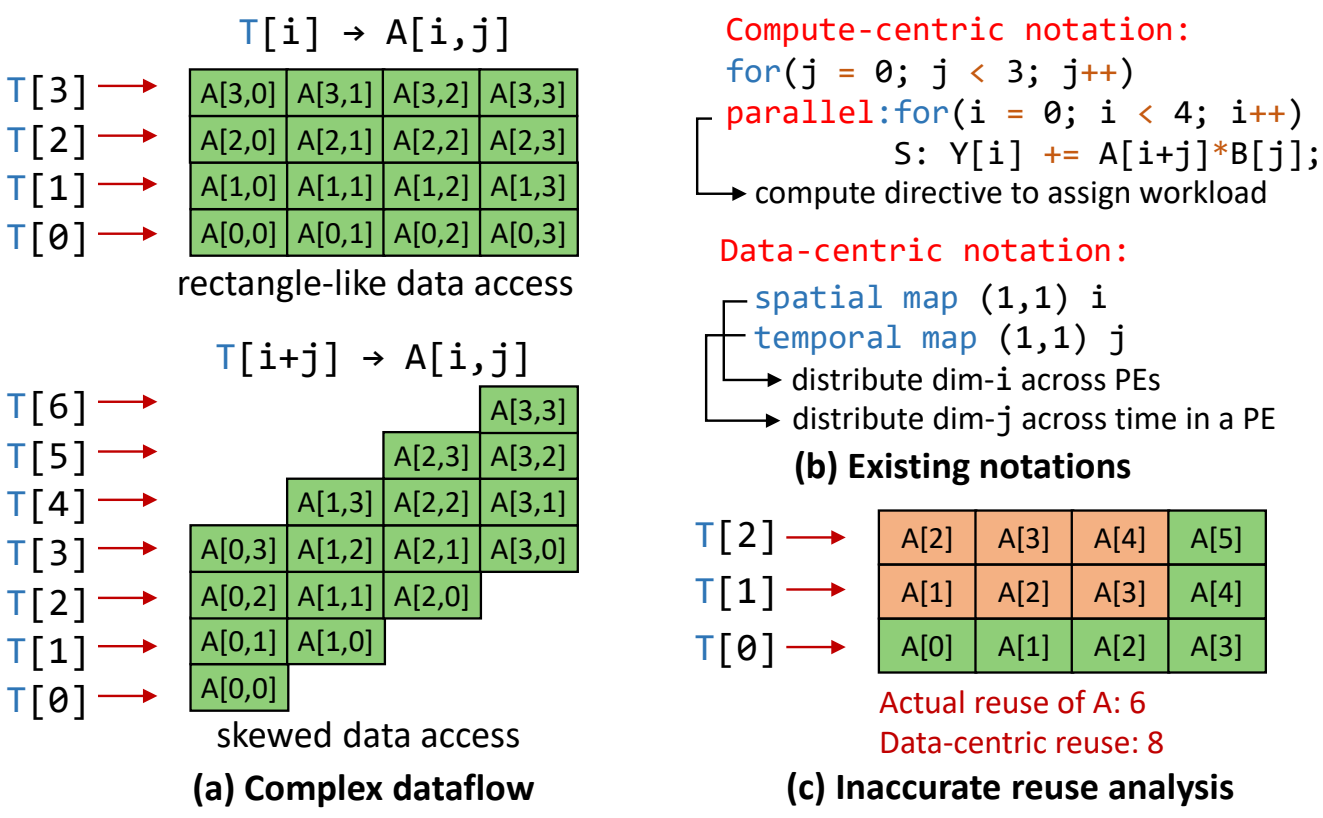

**Fig. 1:** Limitation of compute-centric and data-centric notation.

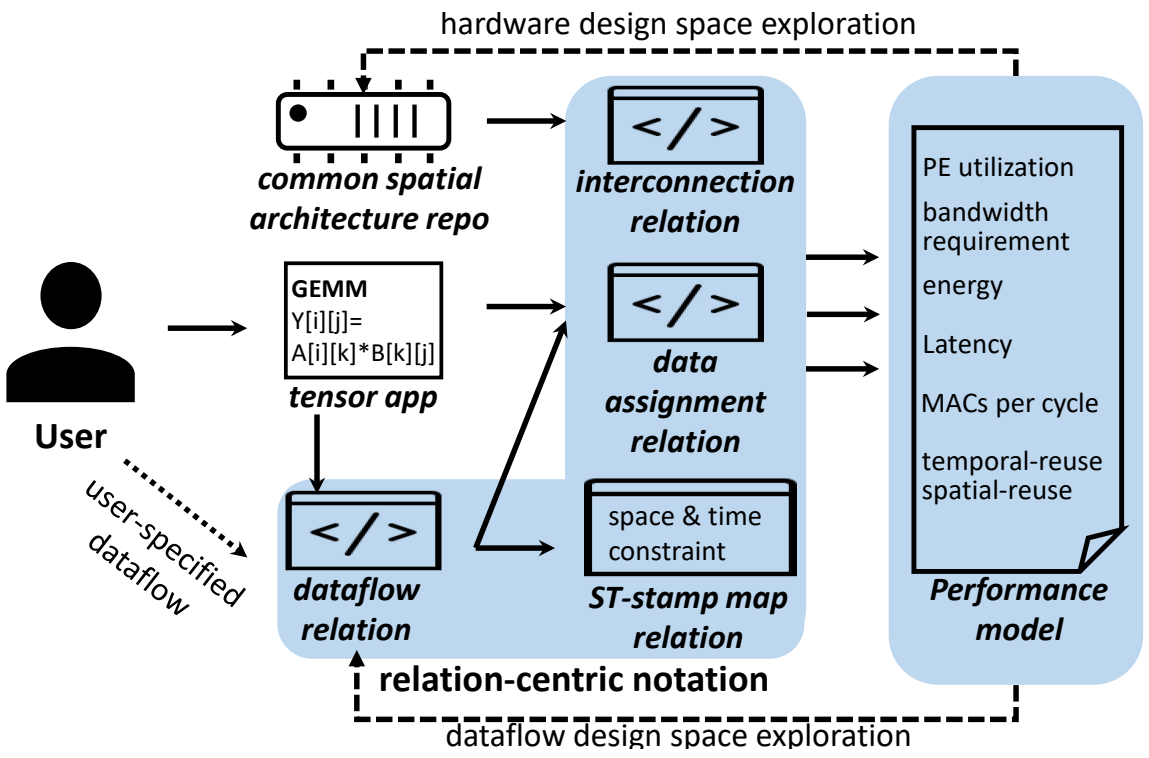

**Fig. 2:** TENET automatic flow.

### *B. Notation Basics*

Here, we introduce some basic terms that are widely used in various compiler frameworks [6, 29, 55] for loop analysis. In this paper, TENET supports tensor applications with perfectly-nest loops and single unconditional statement.

**Iteration domain.** Given a loop nest with one statement $S$, its iteration domain $D_S$ is the set that contains all the loop instances. Each instance can be represented as $S[\vec{n}]$, which is a point in $D_S$. For example, in 1D-CONV operator of Figure 1, the iteration domain is $D_S = \{S[i,j] : 0 \le i < 4, 0 \le j < 3\}$, where $S[i,j]$ is a loop instance and $0 \le i < 4, 0 \le j < 3$ gives the affine constraints.

**Access function.** Given a loop instance, the access function returns the tensor elements accessed by the statement $S$. We use a relation to represent the access function of tensor $F$.

$$A_{S,F} = \{S[\vec{n}] \to F[\vec{f}]\} \tag{1}$$

For example, in 1D-CONV operator of Figure 1, the access function to tensor Y is $\{S[i,j] \to Y[i] : 0 \le i < 4, 0 \le j < 3\}$, which means that the loop instance $S[i,j]$ accesses the tensor element $Y[i]$.

### *C. Limitations of Existing Dataflow Notations*

Table I compares four notations with their features of dataflow expression and performance modeling. For compute-centric notation, the dataflow is specified using loop transformation directives including reorder, blocking, and parallel [39, 56]. The compute-centric notation provides great flexibility for describing how the computation is performed in an imperative programming style. Recently, data-centric notation is proposed, which is specified using data mapping directives including spatial and temporal map, data movement order [24, 25]. Such structural representation enables easy data reuse computation. To use this notation, the users must manually write the directives, which is not straightforward for complex dataflows. Space-time transformation (STT) has been used to map loop instances onto systolic arrays [4, 9, 28]. However, they lack accurate performance models to analyze various hardware metrics, and support for non-systolic array spatial architecture.

Both Timeloop [39] and Interstellar [56] use compute-centric notation. The loop order determines the loop instance execution sequence. However, the workload assignment requires extra directives. For example, in Figure 1(b), the parallelism is specified using `parallel` directive. Data-centric notation explicitly allocates data to PEs with two key primitives, including `Spatial Map` and `Temporal Map`. `Spatial Map` assigns a certain dimension to the PE array, and `Temporal Map` specifies the data movement in a certain dimension across different time-stamps. In Figure 1(b), `spatial map (1,1) i` means distributing the dimension of the output data across different PEs. `temporal map (1,1) j` distributes the data involved in dimension $j$ across different time-stamp within the same PE.

Both compute- and data-centric notations are fundamentally limited in their expressiveness and performance modeling capability. First, both notations fail to cover a complete design space of dataflow. For example, in Figure 1(a), we use $T[t]$ to denote the tensor elements that are processed in cycle $t$. These two notations can only describe dataflows using rectangle-like data access, lacking the support for complex dataflows with skewed data access. Such skewed data access requires the introduction of a new dimension by combing tensor dimensions $i$ and $j$ using affine transformations. More importantly, the limitation in expressiveness cannot be easily remedied by extending the data-centric notation, because it requires great effort to manually transform the original tensor application to explicitly specify the data distribution. Such manual transformation makes it very difficult to estimate the performance metrics, which contradicts with the original intention of the data-centric notation.

Second, from performance modeling perspective, previous compute-centric notation-based models only analyze data reuse opportunities in a coarse-grained manner [39, 56]. For example, Interstellar [56] calculates data reuse using the product of `unroll` factors. The data-centric notation analyzes the hardware performance using MAESTRO model [25]. However, MAESTRO uses simple polynomials to estimate data reuse, which is less precise as it calculates the data movements using a simple polynomial. In Figure 1(c), the actual reuse of tensor A is 6, while the result from MAESTRO is 8. The inaccuracy comes from the fact that the two primitives in Figure 1(b)

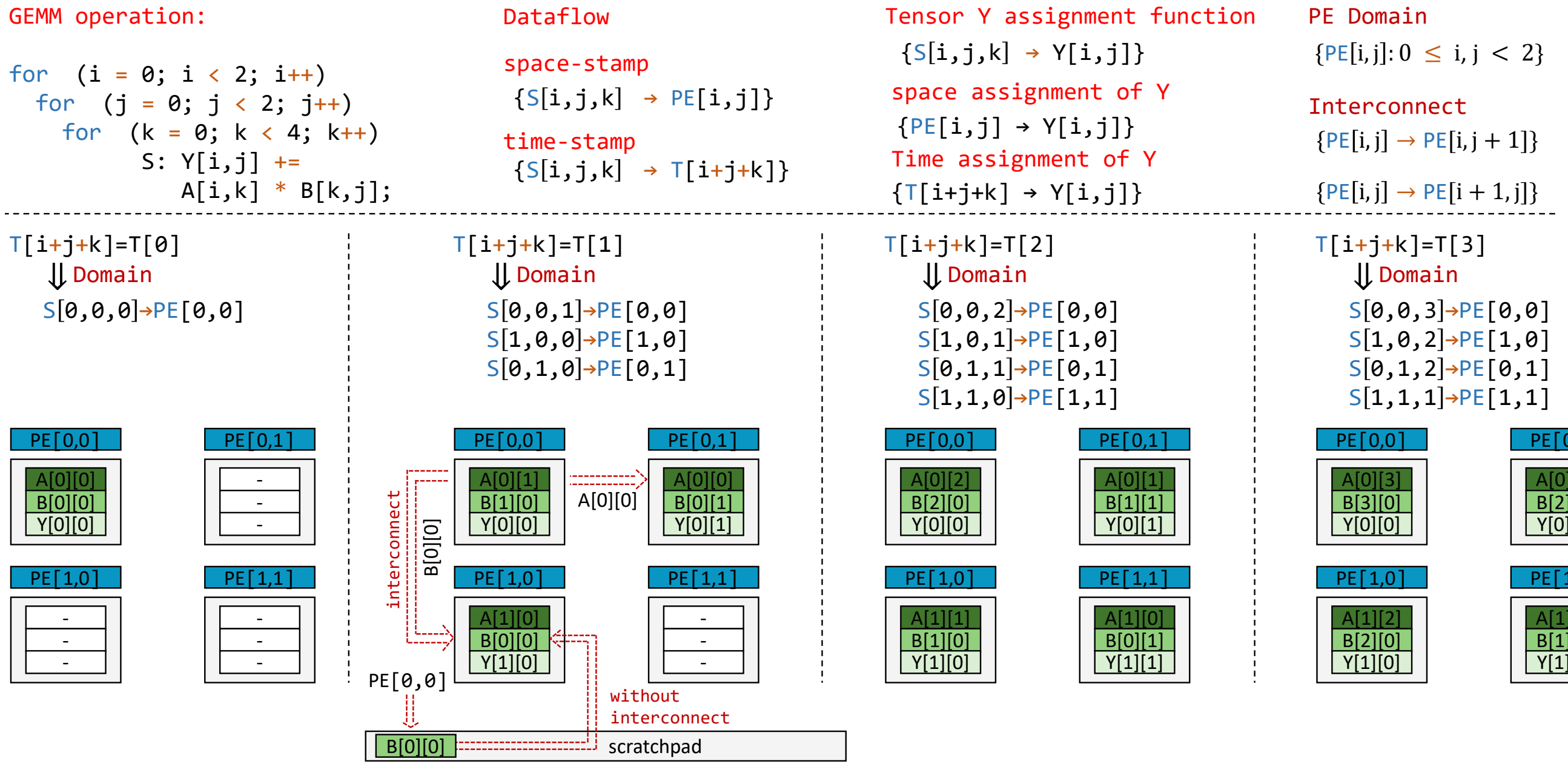

**Fig. 3:** Analyzing the dataflow of matrix multiplication on a $2 \times 2$ PE array using the relation-centric notation.

only describe the data movement of tensor B and Y, without modeling the movement of tensor A.

## III. TENET Overview

TENET is an automatic framework as shown in Figure 2, which takes a tensor operation written in C and hardware specification as inputs. Then, TENET automatically generates the relation-centric notation including dataflow, data assignment, interconnection, and spacetime-stamp map relations. After that, TENET calculates several key performance metrics to guide the dataflow and hardware optimization.

The dataflow relation relates each loop instance to a PE, which can be either generated automatically using design space exploration (DSE) or specified manually by the user. Data assignment relation between tensors and PEs is obtained by combining the access function of the tensor operation and the dataflow relation. The interconnection relation describes how the PEs are connected, which is derived based on the architecture. For a specific dataflow, TENET calculates multiple spacetime-stamp map relations, which will be used for performance metrics estimation. TENET also provides a repository that contains common spatial architectures such as mesh structure, systolic array, reduction tree, which feature with different PE functionalities and PE array topologies.

The expressivity of relation-centric notation can help the accelerator designers to explore the complete design space of the hardware dataflows and find the dataflows that meet their design constraints. Moreover, accelerator designers can rely on TENET to obtain critical performance metrics for a given dataflow including data reuse, PE utilization, and latency.

## IV. Relation-centric Notation

TENET defines four relations, including the mapping of loop instances onto the PE array (Section IV-A), data assignment (Section IV-B), PE interconnection (Section IV-C), mapping between different spacetime-stamps (Section IV-D). Using the relation-centric notation, we can determine exactly where and when each loop instance is executed in the spatial architecture, where and when a tensor element is accessed, and how a tensor element is moved across PEs.

### A. Dataflow Relation

Our dataflow relation applies affine transformation to define the *space-stamp* and *time-stamp* for the execution of tensor applications on a spatial architecture. The space-stamp is a relation that describes the PE coordinates where a loop instance is executed; and the time-stamp is a relation that determines the execution sequence when a loop instance is performed in a PE. Our notation is represented as follows.

*Definition 1:* **Dataflow.** Given a statement S with iteration domain $D_S$ and iteration vector $\vec{n}$, the dataflow is defined as

$$\Theta_{S,D} = \{S[\vec{n}] \rightarrow (PE[\vec{p}] \mid T[\vec{t}])\}, \quad S[\vec{n}] \in Ds \tag{2}$$

$\Theta_{S,D}$ assigns loop instance $S[\vec{n}]$ to a spacetime-stamp, which is a pair of space-stamp ($PE[\vec{p}]$) and time-stamp ($T[\vec{t}]$). The space-stamp gives the coordinates of PE where $S[\vec{n}]$ will be executed, and the time-stamp decides the execution sequence of $S[\vec{n}]$. The sequence is determined by the lexicographical order of two time-stamps. For simplicity, we use $>$ and $<$ to represent the lexicographically larger and smaller, respectively. $\vec{p}$ can be multi-dimensional depending on the PE array dimensions. $\vec{t}$ can also be multi-dimensional as the PE array size can be smaller than the total iteration domain.

Relation-centric notation can express various dataflows by employing affine transformation, as each dimension of the spacetime-stamp can be a linear transformation of multiple loop dimensions. An example of mapping GEMM on a systolic array is demonstrated in Figure 3. We use the following relation-centric notation to describe the dataflow.

$$\Theta_{S,D} = \{S[i,j,k] \to (PE[i,j] \mid T[i+j+k])\}$$

In this example, loop instance $S[i,j,k]$ is executed on $PE[i,j]$, and is assigned a one-dimensional time-stamp $(i+j+k)$. The time-stamp is an affine transformation of loop iterators. Given specific space and time constraints, we can find all the loop instances by solving the constraints. For example, at time-stamp $T[1]$, the set of executed loop instances include

$$\begin{aligned} i+j+k &= 1 \\ constraint: &0 \le i,j < 2, \quad 0 \le k < 4 \\ loop\ instances: &[i,j,k] = [0,0,1],[1,0,0],[0,1,0] \end{aligned}$$

**Dataflow Design Space.** Relation-centric notation is more expressive than compute- and data-centric notations. To make a fair comparison, we assume each PE only has one MAC unit, and the PE array has 2 dimensions. Besides, we also assume that the *size*, *offset* parameters in data-centric notation are set to 1, and the coefficient of affine transformation in relation-centric notation is set to 1. Under these assumptions, the relation-centric notation enlarges the design space from $O(n!\binom{n}{2})$ [33] to $O(2^{n^2})$, where $n$ is the number of loops. The former is the design space of MAESTRO [33], where $n$ primitives can be arranged freely. Each of the primitive can be either *SpatialMap* or *TemporalMap*, and exactly two of them are *SpatialMap*. The later is the number of possible dataflows in relation-centric space. Each dataflow corresponds to an affine transformation, which can be represented using an $n \times n$ transformation matrix filled only with 0 or 1. For example, in GEMM, where $n = 3$, the design space size of MAESTRO is $3! \times 3 = 18$. In contrast, the design space of TENET is $2^9 = 512$, which is $28\times$ larger.

In practice, the loop bound can be much larger than the PE array size. Therefore, we support to use modulus and division operator in the affine transformation (quasi-affine transformation). For example,

$$\begin{aligned} \Theta_{S,D} = \{&S[i,j,k] \to (PE[i \ mod\ 8, j\ mod\ 8] \mid \\ &T[i/8, j/8, (i\ mod\ 8 + j\ mod\ 8 + k)])\} \end{aligned}$$

In this example, the PE array size is $8 \times 8$. As there are 8 indices of dimension $i$ and $j$ being processed in parallel, the time-stamp is 3-dimensional and we use modulus and division operators to represent the execution sequence.

### B. Data Assignment Relation

We use data assignment relations to specify the tensor elements that are accessed by a specific PE at a specific time-stamp in the dataflow. As mentioned in Section II-B, we can use access function to relate the loop instances to its accessed tensor elements. Therefore, the data assignment relation can be formulated as a chain as follows.

*Definition 2:* **Data assignment.** Given a dataflow $\Theta_{S,D}$, the data assignment is defined as

$$A_{D,F} = \Theta_{S,D}^{-1}.A_{S,F} = \{(PE[\vec{p}] \mid T[\vec{t}]) \to F[\vec{f}]\} \quad (3)$$

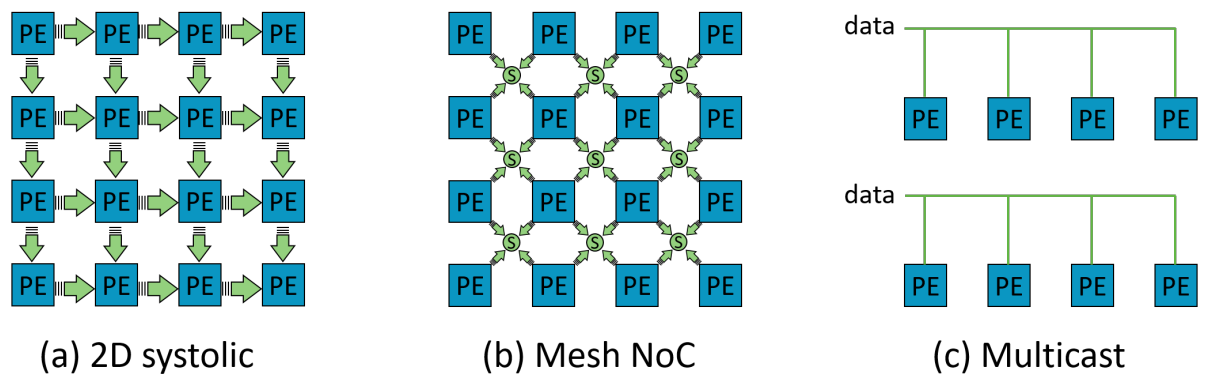

**Fig. 4:** Different interconnect topologies.

In the example of Figure 3, the data assignment of tensor Y can be represented as,

$$A_{D,F_Y} = \{(PE[i,j] \mid T[i+j+k]) \to Y[i,j]\}$$

For this case, we observe that each PE always calculate the same output tensor $(Y)$ at different timestamp, which means tensor $Y[i,j]$ is kept stationary, and iteratively reused at different time-stamps until the computation is finished.

### C. Interconnection Relation

We also specify the PE interconnection using relations, which determines how the tensor elements are moved between different PEs.

*Definition 3:* **PE Interconnection.** Given a PE array, the interconnection is a set, where each element describes a mapping from one PE to another PE.

$$I_{PE_1,PE_2} = \{PE[\vec{p}_1] \to PE[\vec{p}_2]\} \ : c_1, \cdots, c_k \quad (4)$$

where $\vec{p}_1$ and $\vec{p}_2$ denote the coordinate of PEs that are connected, and $c_1, \cdots, c_k$ are the conditions used to specify the topology. The interconnection relation helps to describe the possible data movement in the dataflow. Figure 3 depicts the interconnection specification of a 2D systolic data transfer. The PE is able to reuse the input tensor from adjacent PEs using this interconnection, otherwise, the data must be accessed from the scratchpad.

In this paper, we model three widely used interconnect topologies in Figure 4 as follows.

$$\begin{aligned} \textbf{Interconnection}: &\ \{\mathrm{PE[i,j]} \to \mathrm{PE[i',j']}\} \\ \textbf{2D-systolic}: &\ (i' = i, j' = j+1)\ or\ (i' = i+1, j' = j) \\ \textbf{Mesh}: &\ abs(i'-i) \le 1\ and\ abs(j'-j) \le 1 \\ \textbf{1D-Multicast}: &\ abs(i'-i) \le 3 \ \ (4\ PEs) \end{aligned}$$

The systolic interconnect is widely used in recent GEMM and CONV accelerators like TPU [22]. Mesh NoCs are applied in DySER [19] and Plasticine [42]. In a multicast network, PEs are connected through wires that share the same input entry. Multicast networks are used in Eyeriss [10, 11] and vector dot-product units like Diannao [8]. Different from the first two interconnects, the data reuse of multicast occurs at the same cycle. Details of how interconnection relation affects dataflow modeling will be discussed in Section V-A.

### D. Spacetime-stamp Map Relation

To compute various performance metrics, we need to build relations between different spacetime-stamps. More clearly, by using data assignment and interconnection relations, we can correlate different spacetime-stamps based on the accessed data elements and their movement.

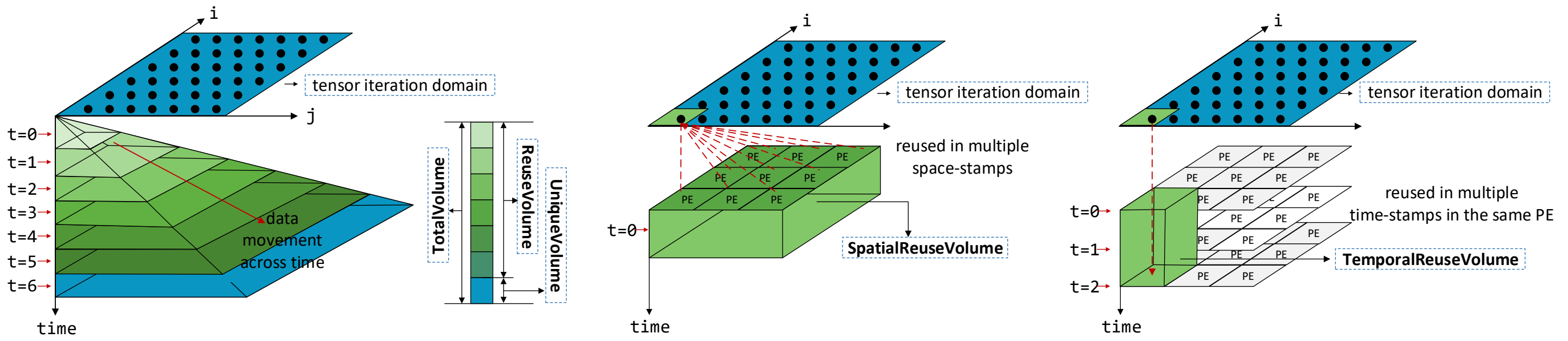

(a) TotalVolume, ReuseVolume, UniqueVolume (b) SpatialReuseVolume (c) TemporalReuseVolume

**Fig. 5:** Volume metrics.

*Definition 4:* **Spacetime-map.** Given a dataflow, the spacetime-map is a relation set that maps a set of spacetime-stamp $D$ to another set of spacetime stamp $D'$. Both $D$ and $D'$ are derived from the given dataflow.

$$M_{D,D'} = \{([PE[\vec{p_1}] \mid T[\vec{t_1}]]) \rightarrow ([PE[\vec{p_2}] \mid T[\vec{t_2}]])\} \quad (5)$$

Given a dataflow, the spacetime-map helps to link different time-stamps and space-stamps so that we can model the hardware behavior in continuous space-stamps and time-stamps. Furthermore, by identifying which data are used by these spacetime-stamps using data assignment relation, we can detect data reuse both spatially and temporally. Assuming the time distance of $D$ and $D'$ is within 1, and the PE specified by $D$ and $D'$ are interconnected, we write three spacetime-stamp maps in the example of Figure 3 as follows.

$$\begin{aligned} &\textbf{map 1.} \quad ([PE[0,0] \mid T[0]) \rightarrow ([PE[0,0] \mid T[1]) \\ &\textbf{map 2.} \quad ([PE[0,0] \mid T[0]) \rightarrow ([PE[0,1] \mid T[1]) \\ &\textbf{map 3.} \quad ([PE[0,0] \mid T[0]) \rightarrow ([PE[1,0] \mid T[1]) \end{aligned} \quad (6)$$

For the **map 1**, by examining the behavior of tensor Y with the assignment relation $A_{D,F_Y}$, we can observe that

$$\begin{aligned} \textbf{map 1.} \quad & ([PE[0,0] \mid T[0]) \rightarrow Y[0,0] \\ & ([PE[0,0] \mid T[1]) \rightarrow Y[0,0] \end{aligned}$$

which means, in this map, tensor element $Y[0,0]$ is reused in the same PE but at different time-stamps. Similarly, we can check the reuse of tensor A and tensor B in the other two maps using $A_{D,F_A}$ and $A_{D,F_B}$, respectively.

$$\begin{aligned} \textbf{map 2.} \quad & ([PE[0,0] \mid T[0]) \rightarrow A[0,0] \\ & ([PE[0,1] \mid T[1]) \rightarrow A[0,0] \\ \textbf{map 3.} \quad & ([PE[0,0] \mid T[0]) \rightarrow B[0,0] \\ & ([PE[1,0] \mid T[1]) \rightarrow B[0,0] \end{aligned}$$

**map 2** and **map 3** describe that tensor element $A[0,0]$ traverses across the PE array horizontally and tensor element $B[0,0]$ traverses vertically, respectively.

In order to estimate various performance metrics such as data reuse, we can apply further restrictions to spacetime-map. For example, on the one hand, by restricting $D$ and $D'$ containing the same PE (e.g. map 1), we can capture the temporal data reuse. On the other hand, by restricting $D$ and $D'$ containing the interconnected PEs (e.g. map 2 and map 3), we can compute spatial data reuse.

**TABLE II:** Details of volume calculation.

| | |
|---|---|
| **TotalVolume** | $TotalVolume = sum(A_{D,F})$ |
| **ReuseVolume** | $ReuseVolume = sum(A_{D,F} \cap M^{-1}_{D,D'}.A_{D,F})$ |
| **UniqueVolume** | $UniqueVolume = TotalVolume - ReuseVolume$ |
| **ReuseFactor** | $ReuseFactor = TotalVolume/UniqueVolume$ |

## V. Performance Model

Based on relation-centric notation, we can compute various performance metrics precisely. The formulation of each metric is naturally modeled as a set operation between relations. In this section, we present how to compute the performance metrics, including data transfer, data reuse, latency, and energy.

### A. Data Reuse and Volume Model

The modeling for hardware metrics starts with defining several data volumes. The value of a specific volume is calculated using sum function (denoted as $sum()$), which counts the cardinality of the set. Table II presents the formulas for these metrics.

**TotalVolume** is the total number of the tensor data accesses through the entire spacetime-stamp. As shown in Figure 5 (a), TotalVolume sums up the volume of all the tensor data accesses that are used across the entire iteration domain. This metric gives the maximum data size that is required to transfer between the PE and the scratchpad. For example, the TotalVolume of tensor A in Figure 3 can be calculated as,

**time-stamp 0. used data** $A[0][0]$
**time-stamp 1. used data** $A[0][1]$, $A[0][0]$ $A[1][0]$
**time-stamp 2. used data** $A[0][2]$, $A[0][1]$, $A[1][1]$, $A[1][0]$
**time-stamp 3. used data** $A[0][3]$, $A[0][2]$, $A[1][2]$, $A[1][1]$
**TotalVolume** $= 1+3+4+4 = 12$

**ReuseVolume** is the number of reused data across multiple spacetime-stamps. For example, in Figure 5 (a), ReuseVolume sums up the volume of data that are overlapped between two adjacent spacetime-stamps. The spacetime-map $M_{D,D'}$ determines the adjacent spacetime-stamps, where the space-stamp map is restricted by the interconnection relation. Only the two stamps meet the constraints in Equation 4 will contribute to data reuse. The time-stamp constraint is defined as a time interval, within which the data reuse can occur. For example, let us assume the interconnect is a systolic topology and the time interval is 1. From time-stamp 1 to time-stamp 3, the ReuseVolume of tensor A in Figure 3 can be calculated as

**time-stamp 1. reused data from time-stamp 0** $A[0][0]$
**time-stamp 2. reused data from time-stamp 1** $A[0][1]$, $A[1][0]$
**time-stamp 3. reused data from time-stamp 2** $A[0][2]$, $A[1][1]$
**ReuseVolume** $= 1+2+2=5$

For multicast interconnection relation, the time interval constraint is set to 0 as data are reused via wires.

**UniqueVolume** is the number of unique tensor data that are accessed. As shown in Figure 5(a), the required tensor data expands as the time-stamp increases. UniqueVolume sums up the volume of "new" data at different spacetime-stamps. A data assigned to a certain spacetime-stamp is considered unique if it cannot be fetched from adjacent spacetime-stamps, where the adjacency is also determined by the spacetime-map (same constraints in $M_{D,D'}$ of ReuseVolume). This metric gives the minimum data size that is required to transfer between the PE and the scratchpad. For example, from time-stamp 0 to time-stamp 3, the UniqueVolume of tensor A in Figure 3 can be calculated as,

**time-stamp 0. new data** $A[0][0]$
**time-stamp 1. new data** $A[0][1]$, $A[1][0]$
**time-stamp 2. new data** $A[0][2]$, $A[1][1]$
**time-stamp 3. new data** $A[0][3]$, $A[1][2]$
**UniqueVolume** $= 1+2+2+2=7$

**ReuseFactor** describes how many times on average a data is reused once it is fetched from scratchpad memory.

**SpatialReuseVolume** is the amount of data reuse across multiple space-stamps. As shown in Figure 5(b), the spatial reuse originates from broadcasting the tensor data to multiple PEs. The SpatialReuseVolume sums up the volume of data with spatial reuse at different space-stamps, where $D$ and $D'$ contain interconnected PEs only.

**TemporalReuseVolume** is the amount of data reuse across multiple time-stamps within the same PE. As shown in Figure 5(c), temporal reuse means that the data is reused across time-stamps. To avoid overlap between SpatialReuseVolume and TemporalReuseVolume, we restrict the TemporalReuseVolume that it only refers to the temporal reuse at the same PE, where $D$ and $D'$ contain the same PE. Clearly, ReuseVolume is the sum of SpatialReuseVolume and TemporalReuseVolume.

### B. Latency and Bandwidth Model

The latency of dataflow consists of communication and computation delay in each PE. We assume buffers, networks and arithmetic units work in a pipelined fashion, and techniques such as double buffering are used to hide latency. Then, the overall latency is just the maximum of communication delay and computation delay.

To model communication delay, we calculate the volume of data that need to be transferred between the on-chip scratchpad and local registers. If a PE can fetch data from itself or its neighboring PEs, then fetching data from the scratchpad can be avoided, which requires more energy and longer latency. Therefore, the volume of data transferred from the scratchpad to registers is estimated by the UniqueVolume of all input tensors. The volume of data sent from output registers to the scratchpad is estimated by the UniqueVolume metric of output tensors. The communication delays are estimated by

$$\begin{aligned} Delay_{read} &= \frac{UniqueVolume(Input)}{bandwidth} \\ Delay_{write} &= \frac{UniqueVolume(Output)}{bandwidth} \end{aligned} \tag{7}$$

Note that the *bandwidth* here is the available scratchpad bandwidth.

The compute delay is estimated by the total number of loop instances and the average number of active PEs.

$$Delay_{compute} = \frac{sum(D_S)}{Util_{PE} \times PE\ size} \tag{8}$$

where $Util_{PE}$ is the average PE utilization across all timestamps.

Using the computation delay, we can also calculate two types of bandwidth requirement, namely, interconnection bandwidth (IBW) and scratchpad bandwidth (SBW). IBW refers to the data communication among PEs, and its is estimated as follows,

$$IBW = \frac{SpatialReuseVolume}{Delay_{compute}} \tag{9}$$

SBW refers to the data transfer between the PE array and the scratchpad, and SBW requirement is estimated by normalizing the UniqueVolume to the computation latency.

$$SBW = \frac{UniqueVolume}{Delay_{compute}} \tag{10}$$

### C. Model Implementation

The performance analysis is written in C++, which leverages the ISL library and Barvinok Library to perform integer set operations [51, 52]. More specifically, we use operators from ISL and Barvinok library to calculate the metrics discussed above. Each relation (e.g., dataflow, interconnection, and data assignment) is implemented using *isl_union_map* structures in ISL. *isl_union_map_reverse* calculates the reverse of an operation (e.g., get $\Theta_{S,D}^{-1}$ from $\Theta_{S,D}$), *isl_union_map_apply_range* composites two relations (e.g., Equation 4). *isl_union_map_card* and *isl_union_pw_qpolynomial_sum* are operators used to summing over time-stamps to calculate metrics such as TotalVolume and UniqueVolume (e.g., Equation 8).

## VI. Evaluation

### A. Experiment Setup

**Benchmarks.** We first evaluate TENET with five important tensor kernels, i.e., GEMM, 2D-CONV, Matrix multiplication chain (MMc), Matricized tensor times Khatri-Rao product (MTTKRP), and Jacobi-2D. GEMM 2D-CONV, and MMc are widely used in deep learning, scientific and engineering computations. MTTKRP tensor operation is the bottleneck operation in tensor factorization (e.g., recommender system).

**TABLE III:** Dataflow notations for various tenso applications.

| Benchmark | Dataflow | Relation-centric | Data-centric (Tp:Temporal, Sp:Spatial) |
|---|---|---|---|
| **GEMM** | (IJ-P \| J,IJK-T)<br>applied in TPU [22] | {S[i,j,k] → PE[i%8,j%8]}<br>{S[i,j,k] → T[fl(i/8),fl(j/8),i%8+j%8+k]} | × |
| | (KJ-P \| K,IJK-T) | {S[i,j,k] → PE[k%8,j%8]}<br>{S[i,j,k] → T[fl(j/8),fl(k/8),i+j%8+k%8]} | × |
| | (IK-P \| K,IJK-T) | {S[i,j,k] → PE[i%8,k%8]}<br>{S[i,j,k] → T[fl(i/8),fl(k/8),j+i%8+k%8]} | × |
| | (K-P \| I,J-T) | {S[i,j,k] → PE[k%64]}<br>{S[i,j,k] → T[fl(k/64),i,j]} | 1. SpMap(1,1) K<br>2. TpMap(1,1) I<br>3. TpMap(1,1) J |
| | (J-P \| I,K-T) | {S[i,j,k] → PE[j%64]}<br>{S[i,j,k] → T[fl(j/64),i,k]} | 1. SpMap(1,1) J<br>2. TpMap(1,1) I<br>3. TpMap(1,1) K |
| **2D-CONV** | (KC-P \| $O_Y$,KC$O_X$-T) | {S[k,c,ox,oy,rx,ry] → PE[k%8,c%8]}<br>{S[k,c,ox,oy,rx,ry] → T[fl(k/8),fl(c/8),oy, k%8+c%8+ox]} | × |
| | (K$O_X$-P \| $O_Y$,K$O_X$C-T) | {S[k,c,ox,oy,rx,ry] → PE[k%8,ox%8]}<br>{S[k,c,ox,oy,rx,ry] → T[fl(k/8),fl(ox/8),oy,k%8+ox%8+c]} | × |
| | (KC-P \| C,K$O_X$-T) | {S[k,c,ox,oy,rx,ry] → PE[k%8,c%8]}<br>{S[k,c,ox,oy,rx,ry] → T[oy, fl(c/8),k%8+ox]} | × |
| | (K-P \| $O_X$,$O_Y$-T) | {S[k,c,ox,oy,rx,ry] → PE[k%64]}<br>{S[k,c,ox,oy,rx,ry] → T[fl(k/64),c,ox,oy]} | 1. SpMap(1,1) K;<br>2. TpMap(1,1) C;<br>3. TpMap(Sz($R_X$),1) X;<br>4. TpMap(Sz($R_Y$),1) Y;<br>5. TpMap(Sz($R_Y$),Sz($R_Y$)) $R_Y$;<br>6. TpMap(Sz($R_X$),Sz($R_X$)) $R_X$; |
| | (C-P \| $O_Y$,$O_X$-T) | {S[k,c,ox,oy,rx,ry] → PE[c%64]}<br>{S[k,c,ox,oy,rx,ry] → T[fl(c/64),k,oy,ox]} | 1. SpMap(1,1) C;<br>2. TpMap(1,1) K;<br>3. TpMap(Sz($R_Y$),1) Y;<br>4. TpMap(Sz($R_X$),1) X;<br>5. TpMap(Sz($R_Y$),Sz($R_Y$)) $R_Y$;<br>6. TpMap(Sz($R_X$),Sz($R_X$)) $R_X$; |
| | ($R_Y O_Y$-P \| $O_Y$,$O_X$-T)<br>Motivated by<br>Eyeriss[10] | {S[k,c,ox,oy,rx,ry] → PE[ry+3*(c%4),oy]}<br>{S[k,c,ox,oy,rx,ry] → T[fl(k/16),fl(c/16),ox]} | 1. TpMap(4,4) C;<br>2. TpMap(16,16) K;<br>3. SpMap(Sz($R_Y$),1) Y;<br>4. TpMap(Sz($R_X$),1) X;<br>5. Cluster(Sz($R_Y$),P);<br>6. TpMap(1,1) C;<br>7. TpMap(1,1) K;<br>8. SpMap(1,1) Y;<br>9. SpMap(1,1) $R_Y$; |
| | ($O_Y O_X$-P \| $O_Y$,$O_X$-T)<br>Motivated by<br>Shi-diannao[15] | {S[k,c,ox,oy,rx,ry] → PE[oy%8,ox%8]}<br>{S[k,c,ox,oy,rx,ry] → T[k,c,fl(oy/8),fl(ox/8)]} | 1. TpMap(1,1) K;<br>2. TpMap(1,1) C;<br>3. SpMap(Sz($R_Y$), 1) Y;<br>4. TpMap(10,8) X;<br>5. TpMap(Sz($R_Y$), Sz($R_Y$)) $R_Y$;<br>6. TpMap(Sz($R_X$), Sz($R_X$)) $R_X$;<br>7. Cluster(8, P);<br>8. SpMap(Sz($R_X$), 1) X; |
| | (KC-P \| $O_Y$,$O_X$-T)<br>Motivated by<br>NVIDIA[38] | {S[k,c,ox,oy,rx,ry] → PE[k%8,c%8]}<br>{S[k,c,ox,oy,rx,ry] → T[fl(k/8),fl(c/8),oy,ox]} | 1. SpMap(1,1) K;<br>2. TpMap(8,8) C;<br>3. TpMap(Sz($R_Y$),Sz($R_Y$)) $R_Y$;<br>4. TpMap(Sz($R_X$),Sz($R_X$)) $R_X$;<br>5. TpMap(Sz($R_Y$),1) Y;<br>6. TpMap(Sz($R_X$),1) X;<br>7. Cluster(8, P);<br>8. SpMap(1,1) C; |
| **MTTKRP** | (IJ-P \| J,IJL-T) | {S[i,j,k,l] → PE[i%8,j%8]}<br>{S[i,j,k,l] → T[k,fl(i/8),fl(j/8),i%8+j%8+l]} | × |
| | (KJ-P \| J,KJL-T) | {S[i,j,k,l] → PE[k%8,j%8]}<br>{S[i,j,k,l] → T[i,fl(k/8),fl(j/8),k%8+j%8+l]} | × |
| | (KL-P \| L,KLJ-T) | {S[i,j,k,l] → PE[k%8,l%8]}<br>{S[i,j,k,l] → T[i,fl(k/8),fl(l/8),k%8+l%8+j]} | × |
| **Jacobi-2D** | (I-P \| I,J-T) | {S[i,j] → PE[i%64]}<br>{S[i,j,k,l] → T[fl(i/64),j]} | × |
| | (IJ-P \| I,J-T) | {S[i,j] → PE[i%8,j%8]}<br>{S[i,j] → T[fl(i/8),fl(j/8)]} | × |
| **MMc** | (IJ-P \| J,IJL-T) | {S[i,j,k,l] → PE[i%8,j%8]}<br>{S[i,j,k,l] → T[k,fl(i/8),fl(j/8),i%8+j%8+l]} | × |
| | (KJ-P \| J,KJL-T) | {S[i,j,k,l] → PE[k%8,j%8]}<br>{S[i,j,k,l] → T[i,fl(k/8),fl(j/8),k%8+j%8+l]} | × |

Jacobi-2D is a two-dimensional stencil operation that is often used in image processing.

$$\textbf{2D-CONV} \quad Y(k,ox,oy) = A(c,ox+rx,oy+ry)B(k,c,rx,ry)$$
$$\textbf{GEMM} \quad Y(i,j) = A(i,k)B(k,j)$$
$$\textbf{MTTKRP} \quad Y(i,j) = A(i,k,l)B(k,j)C(l,j)$$
$$\textbf{MMc} \quad Y(i,j) = A(i,k)B(k,l)C(l,j)$$
$$\textbf{Jacobi-2D} \quad Y(i,j) = ((A(i,j)+A(i-1,j)+A(i,j-1) +A(i+1,j)+A(i,j+1))\ /\ 5$$

We also evaluate four real-world large-scale tensor applications as shown in Table IV. GoogLeNet and MobileNet are the state-of-the-art neural networks in deep learning domain, which require GEMM and 2D-CONV. In Alternating least squares (ALS), MTTKRP is a core operation. Transformer is a well-known network architecture for translation tasks, which features a MMc operation.

**Comparison.** We compare TENET with MAESTRO, which is the state-of-the-art data-centric notation with a comprehensive performance analysis. The compute-centric notation [14, 39, 56] has the same level of expressiveness as data-centric notation but lacks of performance models. More clearly, we compare TENET with MAESTRO using GEMM and 2D-CONV benchmarks, as shown in Table III. We run the open source code of MAESTRO. We configure the two frameworks to use the same parameters (PE Number, Bandwidth, Buffer Size). For the interconnection network, we use a mesh network, since MAESTRO models a hierarchical PE array with the assumption that each PE can communicate with adjacent PEs.

**TABLE IV:** Real-world large-scale tensor applications.

| Application | Domain | Tensor operation | Data size |
|---|---|---|---|
| GoogLeNet[49] | Deep learning | 2D-CONV | 6.7M params<br>3 layer types |
| MobileNet[20] | Deep learning | 2D-CONV | 4.2M params<br>4 layer types |
| ALS[5] | Matrix fabrication | MTTKRP | 480K× 18K× 2K |
| Transformer[50] | NLP | MMc | 512,768,1024 |

### B. Dataflow Comparison

Table III lists 20 popular dataflows that we use to evaluate our framework. The dataflows are named according to the space-stamp and time-stamp in the relation-centric notation, in which **fl()** means floor function and **%** means modulus. For multi-dimensional time-stamp, we only write the innermost two dimensions for simplicity. For example, the $(\mathrm{K-P} \mid \mathrm{O_X, O_Y-T})$ dataflow of 2D-CONV assigns dimension K to the PE array, and the last two dimensions of time-stamp involves $\mathrm{O_X}$ and $\mathrm{O_Y}$, respectively. For the dataflow with affine transformation, we put the transformed dimensions together. For example, in the $(\mathrm{IJ-P} \mid \mathrm{J, IJK-T})$ dataflow of GEMM, the last dimension of time-stamp is given by $(i+j+k)$. We also write the dataflows using data-centric notation if supported.

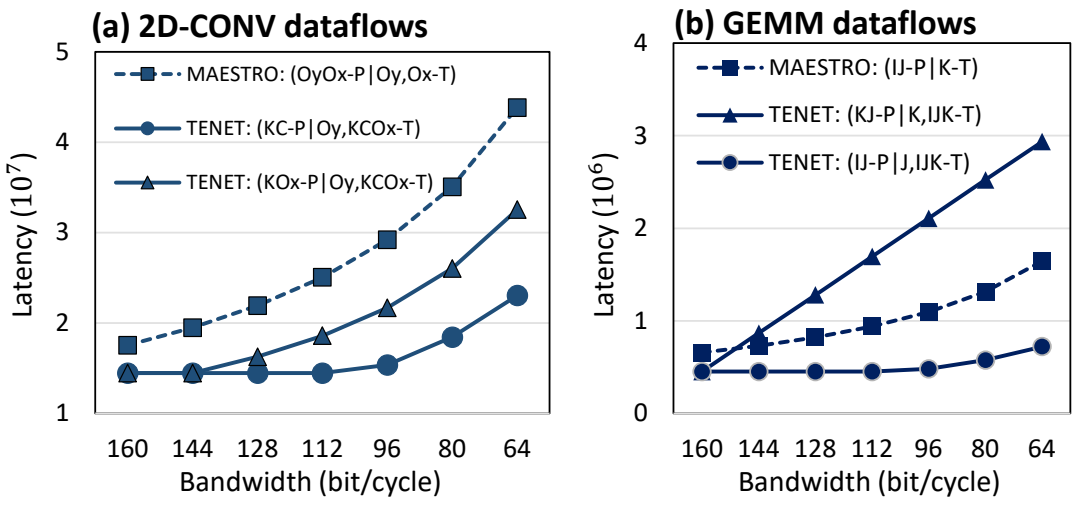

**Fig. 6:** Dataflow comparision for data-centric and relation-centric notations under different bandwidth requirement.

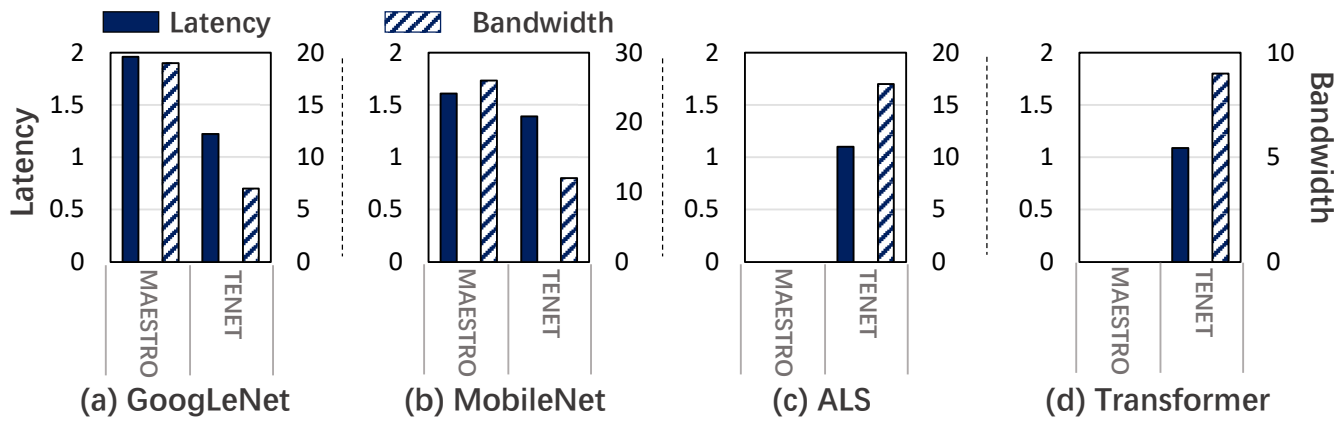

**Fig. 7:** Evaluation on large-scale applications.

Figure 6 presents two use cases of TENET. Figure 6(a) presents the results of 2D-CONV. Two dataflows $(\mathrm{KC-P} \mid \mathrm{O_Y, KCO_X-T})$ $(\mathrm{KO_X-P} \mid \mathrm{O_Y, KO_XC-T})$ cannot be represented in data-centric notation as they require affine transformation, which means relation-centric notation is more expressive than data-centric notation. As a result, relation-centric notation opens up more opportunities for a larger exploration of dataflows, which is essential for designing efficient spatial architectures. In Figure 6, the dataflow depicted in blue lines are the optimal dataflows in data-centric notation. When the bandwidth is 160 bit/cycle, the latency of our dataflow is 17.5% lower than data-centric dataflow. When the bandwidth decreases, the dataflow $(\mathrm{KC-P} \mid \mathrm{O_Y, KCO_X-T})$ achieves up to 47.4% latency reduction compared with the optimal dataflow in data-centric design space. Figure 6 (b) presents the results of GEMM, where two dataflows $(\mathrm{KJ-P} \mid \mathrm{K, IJK-T})$ $(\mathrm{IJ-P} \mid \mathrm{J, IJK-T})$ cannot be represented using data-centric notation. When the bandwidth is 64 bit/cycle, the dataflow $(\mathrm{IJ-P} \mid \mathrm{J, IJK-T})$ achieves up to 77% latency reduction compared with the dataflow in data-centric design space. Overall, TENET achieves 37.4% and 51.4% latency improvement on average compared with the data-centric notation by identifying more sophisticated dataflows for 2D-CONV and GEMM, respectively.

Figure 7 shows the latency and bandwidth requirement of the optimal MAESTRO dataflow and TENET dataflow using the benchmarks in Table IV. The latency is normalized to the ideal latency with theoretical performance (calculated as: # of multipliers $\times$ frequency). The bandwidth is estimated by normalizing the UniqueVolume to the computation latency. MAESTRO model cannot provide the results for the complete ALS and Transform application due to some unsupported operators. We only present TENET results for these two cases. Overall, TENET shows 74% and 22% latency reduction, and reduces the bandwidth requirement by 63% and 54% for GoogleNet and MobileNet, respectively.

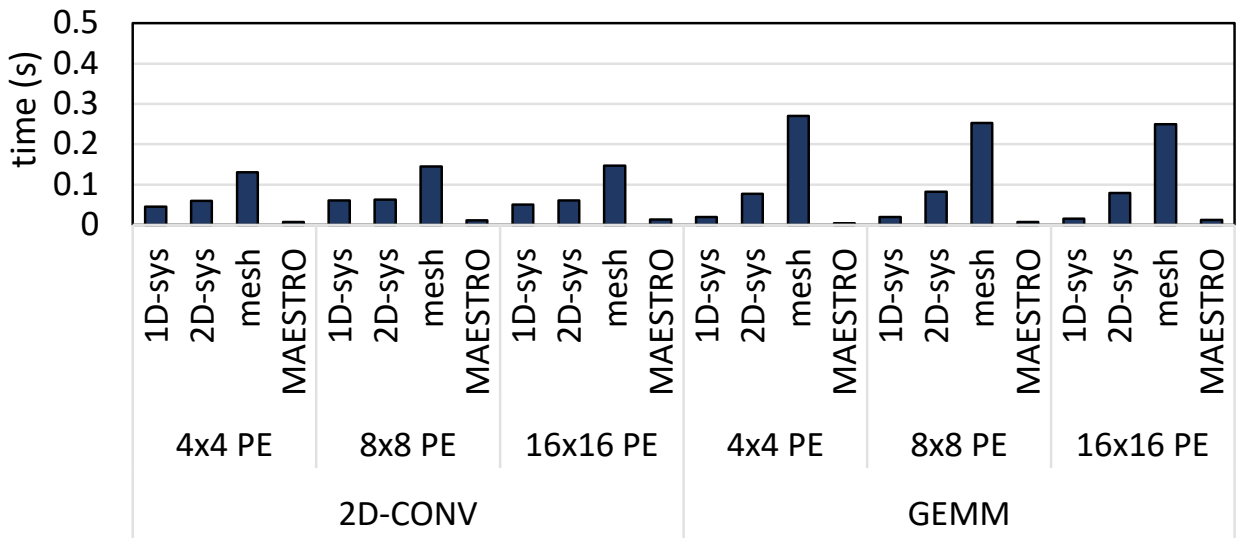

**Fig. 8:** Runtime comparison of TENET and MAESTRO.

Figure 8 compares the execution time of TENET and MASTERO for modeling a single dataflow. The test is conducted on a PC with a 2-core 2.50GHz Intel® Core™ i5-7200U CPU and 8GB memory. On average, the modeling time of a single dataflow is $10^{-2}$ second for MAESTRO, and $10^{-1}$ second for TENET. The difference mainly comes from the fact that TENET models the dataflow as an integer linear programming problem, and considers more architectural details in the evaluation (e.g., interconnection, data assignment). However, MAESTRO calculates metrics using simple polynomials, which is faster but might lead to inaccurate estimation (Section VI-E). We also observe that the modeling time increases for a more complex interconnect, while it is less sensitive to PE array sizes.

**Design space exploration.** The dataflow design space of TENET is huge. We propose to prune the space by restricting the data movement and assignment relations. We first enumerate the possible data movement supported by the PE interconnections, e.g., systolic, multicast for each tensor. Since the data movement is always rectilinear using affine transformation, we then enumerate possible data assignment for the boundary PEs. This together will determine a complete dataflow. Taking 2D-CONV (6 loops) as an example, we only enumerate 12 legal data movements for tensor A and B, respectively. Then, we limit the possible data assignment for the boundary PEs to 180. This results in $12 \times 12 \times 180 = 25920$ different dataflows in total. To explore this design space, TENET takes less than one hour. We leave a more efficient design space exploration as future work.

### C. Performance Metrics Evaluation

Figure 9 shows the evaluation results of five critical metrics: temporal and spatial data reuse of input and output tensors, maximum PE utilization, average PE utilization, and latency. Temporal and spatial data reuse are calculated by normalizing SpatialReuseVolume and TemporalReuseVolume to the number of instances, respectively. In this evaluation, the systolic interconnection topology is applied to all the dataflows.

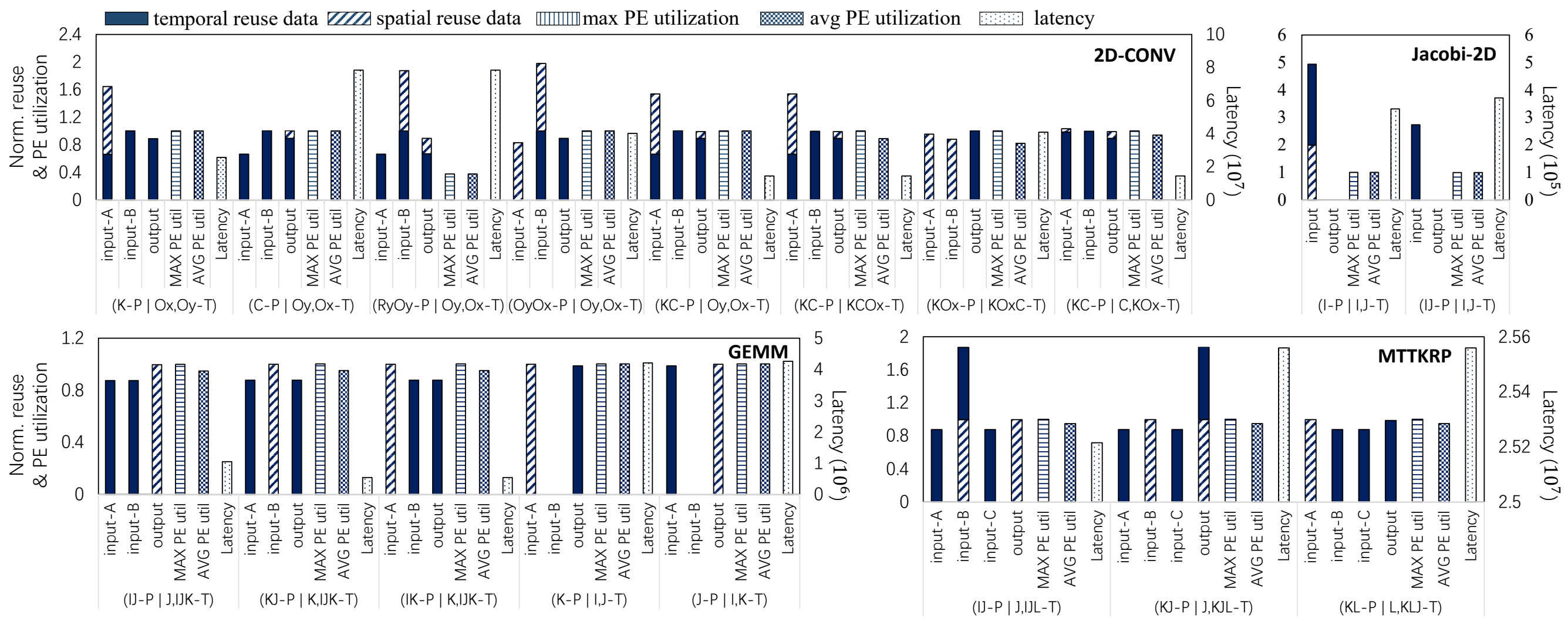

**Fig. 9:** Critical metrics analysis for different tensor applications.

As listed in Table III, we evaluate eight dataflows for 2D-CONV. Different dataflows have high variation in reuse, PE utilization and latency. All these dataflows exhibit temporal and spatial data reuse. However, high data reuse does not necessarily lead to low latency. We observe long latency from the $(\mathrm{R_Y O_Y - P \mid O_Y, O_X - T})$ dataflow. This is because the dimension size of $\mathrm{R_Y}$ cannot match the PE array size, and therefore leads to low PE utilization. We also observe that $(\mathrm{C - P \mid O_Y, O_X - T})$ has high PE utilization but long latency. This is because input-A has low reuse, and therefore loading the input tensor increases the overall latency. Among all 2D-CONV dataflows, the last three have the lowest latency as they utilize all the PEs and have high data reuse for all tensors. For GEMM benchmark, we compare five dataflows. The dataflows that contain two-dimensional space-stamp (i.e., $(\mathrm{IJ - P \mid J, IJK - T})$, $(\mathrm{KJ - P \mid K, IJK - T})$, and $(\mathrm{IK - P \mid K, IJK - T})$) always outperform the ones with one-dimensional space-stamp (i.e., $(\mathrm{K - P \mid I, J - T})$ and $(\mathrm{J - P \mid I, K - T})$) since two-dimensional space-stamp exposes more data reuse opportunities. All these dataflows show high PE utilization while $(\mathrm{IK - P \mid K, IJK - T})$ and $(\mathrm{KJ - P \mid K, IJK - T})$ have the lowest latency, as they have high input reuse on both input tensors. For MTTKRP benchmark, the three dataflows have similar performance, since they all have high PE utilization and large reuse factor on all tensors. Nevertheless, the $(\mathrm{IJ - P \mid J, IJL - T})$ dataflow has higher reuse on input tensors hence outperforms the others.

Noting that a good dataflow needs to provide both high PE utilization and data reuse. Dataflows that have poor performance usually fail in one of these two aspects. The results also demonstrate that our framework is capable of capturing spatial and temporal reuse separately. For example, in the $(\mathrm{IJ - P \mid J, IJK - T})$ dataflow of GEMM, we observe high spatial reuse but little temporal reuse for tensor A and B, since they flow through the PE array. On the other hand, there is high temporal reuse but little spatial reuse for tensor Y, since it is kept stationary in the PE array.

### *D. Bandwidth Analysis*

We also evaluate the bandwidth requirement of three interconnect topologies: 1D-systolic, 2D-systolic, and mesh. When modeling these interconnects, we always assume there exist multicast wires to broadcast the same data to different PEs in the same cycle. In Figure 10, we select five 2D-CONV dataflows that exhibit different types of data reuse. Overall, the three evaluated topologies show similar bandwidth requirement for the same dataflow. This is because tensor applications typically have regular computation and data access patterns, where long-distance communication across PEs is rarely required. Among the four benchmarks, Jacobi-2D is less computation-intensive and requires more scratchpad bandwidth than the others. We also find that the bandwidth requirement comes from different tensors as the dataflow changes. GEMM dataflows illustrate this phenomenon in particular. The dataflow $(\mathrm{IJ - P \mid J, IJK - T})$ maximizes the output reuse by keeping the output stationary in the PE. Therefore, the bandwidth requirement mainly results from the input tensor.

Comparing the three topologies, we also observe some variations. For $(\mathrm{R_Y O_Y - P | O_Y, O_X - T})$ 2D-CONV dataflow and Jacobi-2D dataflow, the mesh topology provides higher interconnection bandwidth and lower scratchpad bandwidth for the input tensor. These two dataflows exhibit diagonal input data reuse, which cannot be supported by both systolic topologies. Since both dataflows can leverage the mesh NoC to increase data reuse, the scratchpad bandwidth requirement is hence reduced. For $(\mathrm{R_Y O_Y - P | O_Y O_X - T})$ 2D-CONV dataflow, we observe that 1D systolic interconnect shows much lower IBW than 2D systolic interconnect, but similar SBW. The saving of IBW comes from the fact that the tensor B is stationary in the PE across different time-stamps. In summary, an interconnection network that connects more PEs does not necessarily reduce scratchpad bandwidth requirement. The design of the interconnection network needs to take the data movement patterns into account.

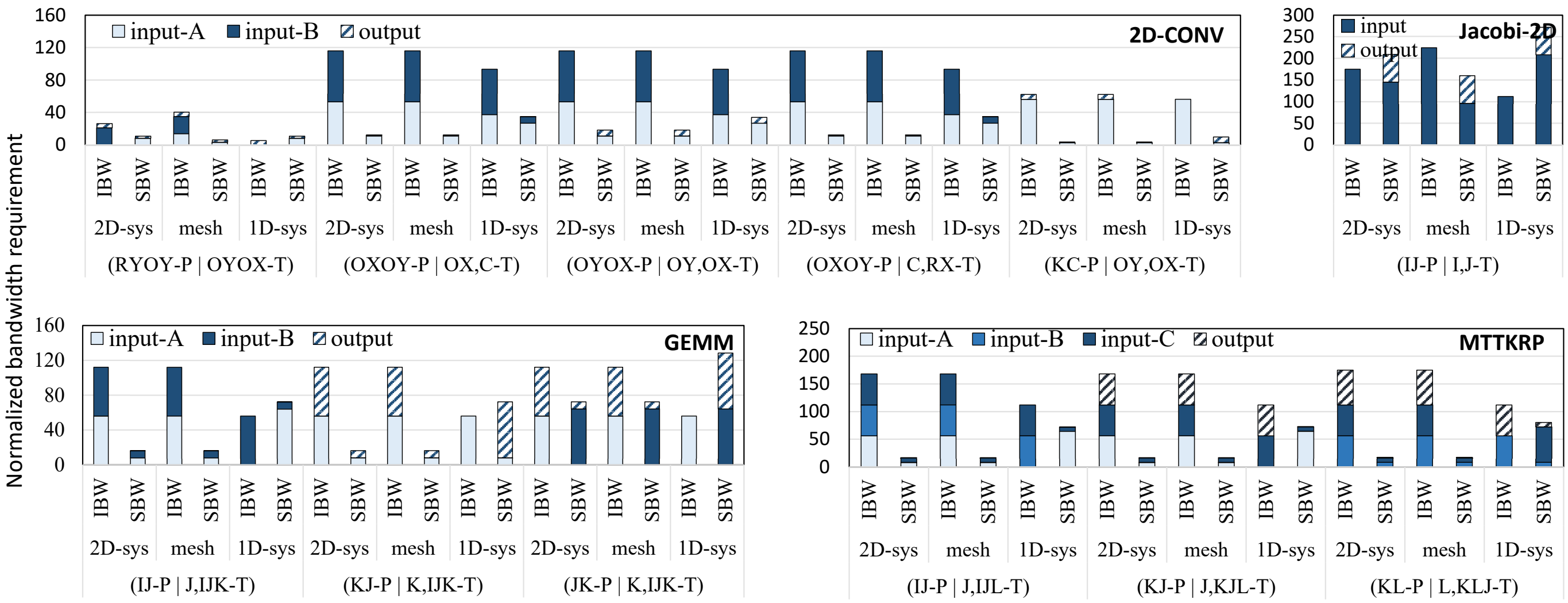

**Fig. 10:** Bandwidth analysis for different interconnection topologies.

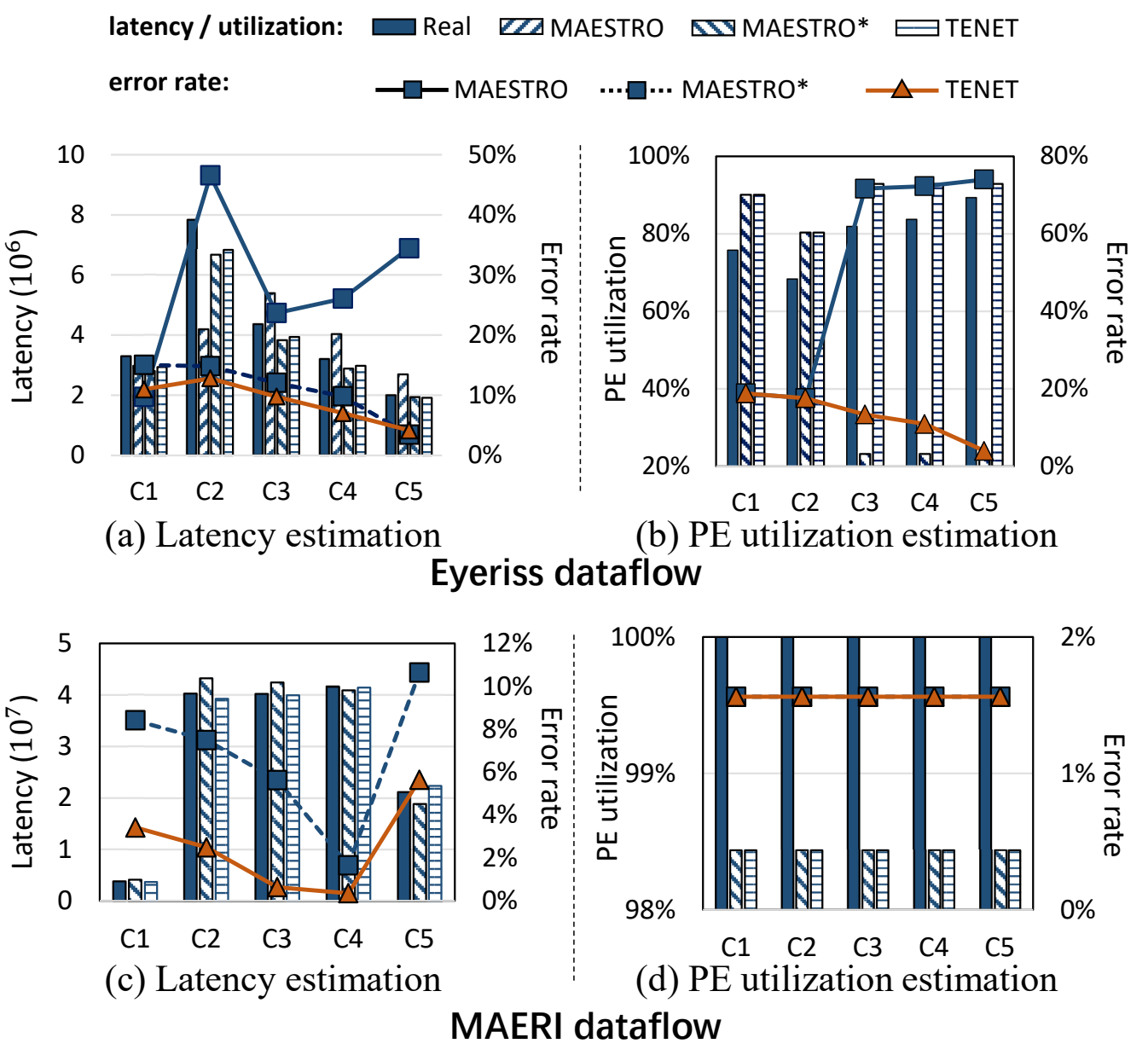

**Fig. 11:** Latency and PE utilization comparison with MAESTRO. (a) and (b) are the results of Eyeriss [10] on AlexNet. (c) and (d) are the results of MAERI [26] on VGG. For VGGNet, C1-C5 means CONV1-1, CONV2-1,...,CONV5-1. According to MAERI [26], the PE utilization is always 100% for dense CNNs.

### *E. Metrics Estimation Comparison*

In this section, we compare TENET with MAESTRO on the accuracy of latency, PE utilization, and reuse estimation. We show that our proposed model improves the metric estimation accuracy. Following previous deep learning accelerators, in 2D-CONV, we term tensor A as input, tensor B as filter, dimension K as output channel, dimension C as input channel.

**Latency Comparison.** We compare the latency calculated by our work and MAESTRO using the row-stationary dataflow proposed by Eyeriss [10] for AlexNet, and using the dataflow proposed by MAERI [26] for VGG. We use the reported latency and PE utilization in Eyeriss and MAERI as the golden result. For MAESTRO implementation, we use the same input files as in [24]. Figure 11 shows the comparison results. Overall, our work improves the latency estimation accuracy from 71.9% to 89.6% for Eyeriss, and from 92.3% to 96.3% for MAERI.

The improvement is partially due to TENET's capability of modeling complex dataflow using affine transformation. In Alexnet, the size of filter $B(k,c,rx,ry)$ varies from $(96,3,11,11)$ to $(384,256,3,3)$. To increase the PE utilization, Eyeriss distributes the dimension $ry$ and $c$ to 12 PEs in one column. Relation-centric notation supports this feature by assigning a space-stamp with the affine transformation $(ry+3*(c\%4))$ to the column of the PE array. However, MAESTRO only models the case when $c=0$ due to the limitation of data-centric notation. Such limitation explains the large error in the last three layers where $ry=3$. We also compare with another Eyeriss dataflow using data-centric notation (denoted as MAESTRO*), which is different from the notation in [24]. In MAESTRO*, the `cluster` primitives are used to merge the multiple channel dimensions so that PEs can be better utilized. The accuracy of MAESTRO* is 89.0%.

MAERI features a reconfigurable reduction tree design, which can be configured to enable multiple vector dot-products with different sizes. For the reduction tree dataflow, only multipliers are considered as PEs and PEs are connected via multicast interconnection. TENET applies affine transformation to denote MAERI dataflow as it assigns multiple dimensions to the 1D PE array. In the implementation of MAESTRO, we also manually transform two dimensions of the filter and the input to fit the 1D PE array of MAERI. TENET improves the accuracy thanks to more accurate reuse analysis.

**PE Utilization Comparison.** TENET estimates the PE utilization accurately, which is another key factor that affects the latency, as shown in Equation 8. We also compare the average PE utilization rate with MAESTRO. Since Eyeriss did not report this metric directly, we approximate the ground-truth value using $\frac{\text{MACNum}}{\text{PENum}\times\text{Latency}}$ which is obtained from Eyeriss. As shown in Figure 11 (b), both frameworks predict the same at the first two layers, but our work makes much more accurate predictions in the following three layers. Unlike previous approaches that estimate PE utilization via a polynomial composed of PE array size and total computation size, TENET models the PE utilization by going through all time-stamps to check whether a PE is assigned.

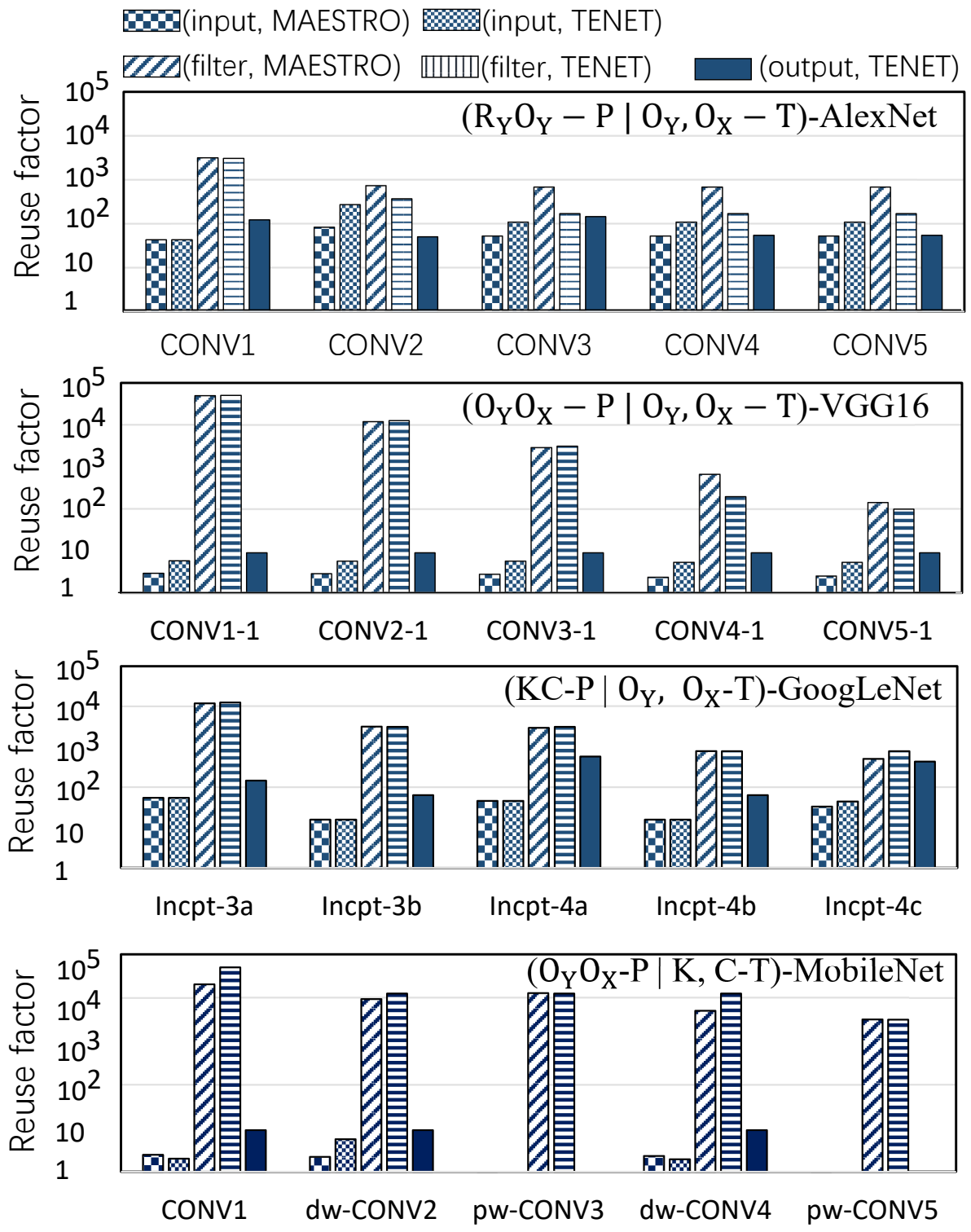

**Fig. 12:** Data reuse comparison with MAESTRO.

**Reuse Comparison.** TENET can precisely calculate the ReuseFactor that affects the latency for data transfer. We evaluate the reuse analysis of MAESTRO and our work on four state-of-the-art DNN models: AlexNet, VGG16, GoogleNET, and MobilleNet. We select different dataflows for each model. For AlexNet, we use the row-stationary dataflow on a $12 \times 14$ PE array, motivated by Eyeriss [10]. For VGG16, we use the output stationary dataflow on a $8 \times 8$ PE array, motivated by Shidiannao [15]. We do not normalize the dataflows to be executed on a $8 \times 8$ PE array here because our goal is to accurately capture the characteristics of original dataflows instead of comparing between them.

The first case is the CONV3 layer of AlexNet. In this layer, Eyeriss maps $R_Y$ and C dimensions into dimension 1 of the PE array to make full use of PEs, and $O_Y$ into dimension 2 of the PE array. The filter array has a spatial reuse factor of 13 (size of $O_Y$) since it is reused across PEs horizontally. Besides, the filter array stays stationary in the innermost dimension of time-stamp ($O_X$), which contributes a temporal reuse factor of 13 (size of $O_X$). Therefore the reuse factor should be $13 \times 13 = 169$, which is accurately calculated by our work (169) compared to MAESTRO (676). For output array, since all PEs in a vertical line shares the same output data, it has a spatial reuse factor of 12 (size of dimension 1 of PE array). Moreover, there is a temporal reuse factor of 12 since each PE processes 3 filter width ($R_X$) and 4 input channels (C) continuously, leading to a total reuse factor of $12 \times 12 = 144$, which is also accurately calculated by our work. On the other hand, MAESTRO reports no reuse for the output array in all circumstances, which is likely because the dimensions of output array $(O_X, O_Y)$ are not explicitly specified by the primitives.

Similar inaccuracy is observed on VGG16, GoogleNet, and MobileNet as shown in Figure 12. For example, the filter reuse in inception-4a calculated by TENET is 3136 while it is 2916 from MAESTRO. The actual filter reuse depends on the input image size, which is $56 \times 56 = 3136$. In MobileNet, there are two special convolutional layers, namely depthwise convolution (dw-CONV) and pointwise convolution (pw-CONV). In the depthwise convolution, results from different input channels are directly stored as the outputs without accumulation. Therefore, this layer shows lower input data reuse. The pointwise convolution uses $1 \times 1$ filter size, which leads to no reuse for input data. MAESTRO is less accurate in data reuse estimation for three reasons. First, the simple formulas used by MAESTRO require all data dimensions being explicitly specified by the primitives, and hence no reuse is reported for output array in all cases. Second, MAESTRO only analyzes data reuse for the innermost temporal and spatial dimensions, while our work can analyze data reuse for multiple time dimensions. Third, MAESTRO does not support mapping multiple data dimensions on a single PE dimension, which is a technique applied by Eyeriss to fill the PE array.

## VII. Related Works

**Notations for expressing the dataflow.** Compute-centric notation is widely adopted as it can directly represents dataflow in high-level languages using directives [39, 56]. In [14], dataflow is notated using two hyperplanes with polyhedral dependency graph. Timeloop [39] describes the design space using a concise and unified loop representation with mapping directives. The mapping directives consist of memory constraints and PE workload assignment. Interstellar [56] represents the hardware dataflow using Halide's scheduling language [45] for design exploration. Interstellar extends Halide with additional control directives, e.g., `loop blocking` and `resource allocation`, for specifying the hardware features. Recently, Kwon *et al.* [24] propose a data-centric notation that used `spatial map` and `temporal map` to specify the dataflow. Space-time transformation theory is widely used for systolic architectures [4, 9, 13, 27, 28, 35]. There are also studies mainly focusing on the arithmetic properties, such as array shape transformation, partitioning of the space coordinate for systolic arrays [4, 9, 28, 35]. Recent efforts attempt to apply this theory for automatic FPGA code generation [13, 27]. PolySA [13] and AutoSA [54] compile applications into polyhedral IR and map them on systolic arrays. SuSy [27] combines polyhedral with hardware optimization primitives to generate high-performance systolic array on FPGAs.

**Modeling and optimization.** Performance modeling can provide general guidelines and insights for optimizing the dataflow. Prior performance models mainly focus on the DNN applications on spatial architectures [11, 25, 32, 56]. Eyeriss-v2 [11] analyzes data reuse with different tensor shapes and sizes. However, it lacks a performance model to evaluate the energy and performance. MAESTRO [25] applies data-centric notation and analyzes data reuse, energy, and latency. Interstellar [56] explores the design space using loop operators like loop split and loop reorder. There are also prior works aiming at modeling the spatial architecture for general applications [29, 37, 41]. Nowatzki *et al.* [37] proposed a constraint-centric scheduling algorithm that determines the placement and routing of reconfigurable architectures. [41] represents the dataflow on the computation graph with edge-centric approach for application mapping. Recently, Jia et al. proposed TensorLib that built parameterized hardware module templates based on relation-centric notation [21]. By feeding the dataflow found by Tenet to TensorLib, we can automatically generate the dataflow hardware written in Chisel [3].

## VIII. Conclusion

In this work, we propose a relation-centric notation that represents dataflow, data assignment and interconnection uniformly as relations. The relation-centric notation supports a larger dataflow design space and more tensor benchmarks compared to prior notations. We also present an analytical framework to accurately estimate data reuse, bandwidth requirement, latency and energy. We evaluate our framework on four tensor benchmarks and results show that the relation-centric notation can find more sophisticated dataflows with lower latency compared to the state-of-the-art data-centric notation. TENET achieves 37% to 51% latency reduction compared to the state-of-the-art techniques.

## Acknowledgements

This work was supported in part by the Beijing Natural Science Foundation (No. JQ19014) and in part by the Beijing Academy of Artificial Intelligence (BAAI). We would like to thank Tushar Krishna for sharing the source code of MASTERO.

## References

[1] M. Abadi, P. Barham, J. Chen, Z. Chen, A. Davis, J. Dean, M. Devin, S. Ghemawat, G. Irving, M. Isard, M. Kudlur, J. Levenberg, R. Monga, S. Moore, D. Murray, B. Steiner, P. Tucker, V. Vasudevan, P. Warden, M. Wicke, Y. Yu, and X. Zheng, "Tensorflow: A system for large-scale machine learning," in *Symposium on Operating Systems Design and Implementation*, 2016.

[2] A. Anandkumar, R. Ge, D. Hsu, S. M. Kakade, and M. Telgarsky, "Tensor decompositions for learning latent variable models," *The Journal of Machine Learning Research*, 2014.

[3] J. Bachrach, H. Vo, B. Richards, Y. Lee, A. Waterman, R. Avižienis, J. Wawrzynek, and K. Asanović, "Chisel: constructing hardware in a scala embedded language," in *Proceedings of the Design Automation Conference*, 2012.

[4] D. G. Baltus and J. Allen, "Efficient exploration of nonuniform space-time transformations for optimal systolic array synthesis," in *Proceedings of International Conference on Application Specific Array Processors*. IEEE, 1993.

[5] J. Bennett and S. Lanning, "The netflix prize," in *Proceedings of KDD cup and workshop*, 2007.

[6] U. Bondhugula, A. Hartono, J. Ramanujam, and P. Sadayappan, "A practical automatic polyhedral parallelizer and locality optimizer," in *Proceedings of the 29th ACM SIGPLAN Conference on Programming Language Design and Implementation*, 2008.

[7] D. Burger, S. W. Keckler, K. S. McKinley, M. Dahlin, L. K. John, C. Lin, C. R. Moore, J. Burrill, R. G. McDonald, and W. Yoder, "Scaling to the end of silicon with edge architectures," *Computer*, 2004.

[8] T. Chen, Z. Du, N. Sun, J. Wang, C. Wu, Y. Chen, and O. Temam, "Diannao: A small-footprint high-throughput accelerator for ubiquitous machine-learning," *ACM Sigplan Notices*, 2014.

[9] Y.-K. Chen and S.-Y. Kung, "A systolic design methodology with application to full-search block-matching architectures," *Journal of VLSI signal processing systems for signal, image and video technology*, 1998.

[10] Y.-H. Chen, J. Emer, and V. Sze, "Eyeriss: A spatial architecture for energy-efficient dataflow for convolutional neural networks," in *ACM SIGARCH Computer Architecture News*, 2016.

[11] Y.-H. Chen, T.-J. Yang, J. Emer, and V. Sze, "Eyeriss v2: A flexible accelerator for emerging deep neural networks on mobile devices," *IEEE Journal on Emerging and Selected Topics in Circuits and Systems*, 2019.

[12] Y. Chen, T. Luo, S. Liu, S. Zhang, L. He, J. Wang, L. Li, T. Chen, Z. Xu, N. Sun *et al.*, "Dadiannao: A machine-learning supercomputer," in *Proceedings of the 47th Annual IEEE/ACM International Symposium on Microarchitecture*, 2014.

[13] J. Cong and J. Wang, "PolySA: Polyhedral-Based Systolic Array Auto-Compilation," in *Proceedings of the International Conference on Computer-Aided Design*, 2018.

[14] S. Dave, A. Shrivastava, Y. Kim, S. Avancha, and K. Lee, "dMazeRunner: Optimizing Convolutions on Dataflow Accelerators," in *IEEE International Conference on Acoustics, Speech and Signal Processing*, 2020.

[15] Z. Du, R. Fasthuber, T. Chen, P. Ienne, L. Li, T. Luo, X. Feng, Y. Chen, and O. Temam, "Shidiannao: Shifting vision processing closer to the sensor," in *ACM SIGARCH Computer Architecture News*, 2015.

[16] D. M. Dunlavy, T. G. Kolda, and W. P. Kegelmeyer, "Multilinear algebra for analyzing data with multiple linkages," in *Graph algorithms in the language of linear algebra*, 2011.

[17] H. Esmaeilzadeh, A. Sampson, L. Ceze, and D. Burger, "Neural acceleration for general-purpose approximate programs," in *45th Annual IEEE/ACM International Symposium on Microarchitecture*, 2012.

[18] J. Fowers, K. Ovtcharov, M. Papamichael, T. Massengill, M. Liu, D. Lo, S. Alkalay, M. Haselman, L. Adams, M. Ghandi *et al.*, "A configurable cloud-scale dnn processor for real-time ai," in *2018 ACM/IEEE 45th Annual International Symposium on Computer Architecture*. IEEE, 2018.

[19] V. Govindaraju, C.-H. Ho, T. Nowatzki, J. Chhugani, N. Satish, K. Sankaralingam, and C. Kim, "Dyser: Unifying functionality and parallelism specialization for energy-efficient computing," *IEEE Micro*, 2012.

[20] A. G. Howard, M. Zhu, B. Chen, D. Kalenichenko, W. Wang, T. Weyand, M. Andreetto, and H. Adam, "Mobilenets: Efficient convolutional neural networks for mobile vision applications," in *IEEE Conference on Computer Vision and Pattern Recognition*, 2017.

[21] L. Jia, Z. Luo, L. Lu, and Y. Liang, "TensorLib: A Spatial Accelerator Generation Framework for Tensor Algebra," *arXiv preprint arXiv:2104.12339*, 2021.

[22] N. P. Jouppi, C. Young, N. Patil, D. Patterson, G. Agrawal, R. Bajwa, S. Bates, S. Bhatia, N. Boden, A. Borchers *et al.*,

"In-datacenter performance analysis of a tensor processing unit," in *Proceedings of the 44th Annual International Symposium on Computer Architecture*, 2017.

[23] T. G. Kolda and B. W. Bader, "Tensor decompositions and applications," *SIAM review*, 2009.

[24] H. Kwon, P. Chatarasi, M. Pellauer, A. Parashar, V. Sarkar, and T. Krishna, "Understanding reuse, performance, and hardware cost of dnn dataflow: A data-centric approach," in *Proceedings of the 52nd Annual IEEE/ACM International Symposium on Microarchitecture*, 2019.

[25] H. Kwon, P. Chatarasi, V. Sarkar, T. Krishna, M. Pellauer, and A. Parashar, "Maestro: A data-centric approach to understand reuse, performance, and hardware cost of dnn mappings," *IEEE Micro*, 2020.

[26] H. Kwon, A. Samajdar, and T. Krishna, "Maeri: Enabling flexible dataflow mapping over dnn accelerators via reconfigurable interconnects," in *ACM SIGPLAN Notices*, 2018.

[27] Y.-H. Lai, H. Rong, S. Zheng, W. Zhang, X. Cui, Y. Jia, J. Wang, B. Sullivan, Z. Zhang, Y. Liang *et al.*, "Susy: A programming model for productive construction of high-performance systolic arrays on fpgas," in *International Conference on Computer Aided Design*, 2020.

[28] M. S. Lam, *A systolic array optimizing compiler*. Springer Science & Business Media, 2012, vol. 64.

[29] D. Liu, S. Yin, L. Liu, and S. Wei, "Polyhedral model based mapping optimization of loop nests for CGRAs," in *Proceedings of the 50th Annual Design Automation Conference*, 2013.

[30] D. Liu, T. Chen, S. Liu, J. Zhou, S. Zhou, O. Teman, X. Feng, X. Zhou, and Y. Chen, "Pudiannao: A polyvalent machine learning accelerator," in *ACM SIGARCH Computer Architecture News*, 2015.

[31] S. Liu, Z. Du, J. Tao, D. Han, T. Luo, Y. Xie, Y. Chen, and T. Chen, "Cambricon: An instruction set architecture for neural networks," in *2016 ACM/IEEE 43rd Annual International Symposium on Computer Architecture*. IEEE, 2016.

[32] W. Lu, G. Yan, J. Li, S. Gong, Y. Han, and X. Li, "Flexflow: A flexible dataflow accelerator architecture for convolutional neural networks," in *2017 IEEE International Symposium on High Performance Computer Architecture*, 2017.

[33] MAESTRO, https://github.com/maestro-project/maestro, 2020.

[34] J. McAuley and J. Leskovec, "Hidden factors and hidden topics: understanding rating dimensions with review text," in *the conference on Recommender systems*, 2013.

[35] I. Milovanović, E. Milovanović, and M. Bekakos, "Synthesis of a unidirectional systolic array for matrix–vector multiplication," *Mathematical and computer modelling*, 2006.

[36] T. Nowatzki, V. Gangadhar, N. Ardalani, and K. Sankaralingam, "Stream-dataflow acceleration," in *ACM/IEEE 44th Annual International Symposium on Computer Architecture*, 2017.

[37] T. Nowatzki, M. Sartin-Tarm, L. De Carli, K. Sankaralingam, C. Estan, and B. Robatmili, "A general constraint-centric scheduling framework for spatial architectures," *ACM SIGPLAN Notices*, 2013.

[38] NVIDIA, http://nvdla.org/, 2020.

[39] A. Parashar, P. Raina, Y. S. Shao, Y. Chen, V. A. Ying, A. Mukkara, R. Venkatesan, B. Khailany, S. W. Keckler, and J. Emer, "Timeloop: A systematic approach to dnn accelerator evaluation," in *2019 IEEE International Symposium on Performance Analysis of Systems and Software*, 2019.

[40] A. Parashar, M. Pellauer, M. Adler, B. Ahsan, N. Crago, D. Lustig, V. Pavlov, A. Zhai, M. Gambhir, A. Jaleel *et al.*, "Triggered instructions: a control paradigm for spatially-programmed architectures," *ACM SIGARCH Computer Architecture News*, 2013.

[41] H. Park, K. Fan, S. A. Mahlke, T. Oh, H. Kim, and H.-s. Kim, "Edge-centric modulo scheduling for coarse-grained reconfigurable architectures," in *Proceedings of the 17th international conference on Parallel architectures and compilation techniques*, 2008.

[42] R. Prabhakar, Y. Zhang, D. Koeplinger, M. Feldman, T. Zhao, S. Hadjis, A. Pedram, C. Kozyrakis, and K. Olukotun, "Plasticine: A reconfigurable architecture for parallel patterns," in *ACM/IEEE 44th Annual International Symposium on Computer Architecture*, 2017.

[43] W. Qadeer, R. Hameed, O. Shacham, P. Venkatesan, C. Kozyrakis, and M. A. Horowitz, "Convolution engine: balancing efficiency & flexibility in specialized computing," in *Proceedings of the 40th Annual International Symposium on Computer Architecture*, 2013.

[44] X. Qingcheng, Z. Size, W. Bingzhe, X. Pengcheng, Q. Xuehai, and L. Yun, "HASCO: Towards Agile HArdware and Software CO-design for Tensor Computation," in *2021 ACM/IEEE 48th Annual International Symposium on Computer Architecture*, 2021.

[45] J. Ragan-Kelley, C. Barnes, A. Adams, S. Paris, F. Durand, and S. Amarasinghe, "Halide: a language and compiler for optimizing parallelism, locality, and recomputation in image processing pipelines," *Acm Sigplan Notices*, 2013.

[46] F. Sadi, J. Sweeney, T. M. Low, J. C. Hoe, L. Pileggi, and F. Franchetti, "Efficient spmv operation for large and highly sparse matrices using scalable multi-way merge parallelization," in *Proceedings of the 52nd Annual IEEE/ACM International Symposium on Microarchitecture*, 2019.

[47] S. Smith and G. Karypis, "Tensor-matrix products with a compressed sparse tensor," in *Proceedings of the Workshop on Irregular Applications: Architectures and Algorithms*, 2015.

[48] S. Swanson, K. Michelson, A. Schwerin, and M. Oskin, "Wavescalar," in *Proceedings. 36th Annual IEEE/ACM International Symposium on Microarchitecture*, 2003.

[49] C. Szegedy, W. Liu, Y. Jia, P. Sermanet, S. Reed, D. Anguelov, D. Erhan, V. Vanhoucke, and A. Rabinovich, "Going deeper with convolutions," in *Proceedings of the IEEE conference on computer vision and pattern recognition*, 2015.

[50] A. Vaswani, N. Shazeer, N. Parmar, J. Uszkoreit, L. Jones, A. N. Gomez, Ł. Kaiser, and I. Polosukhin, "Attention is all you need," *NeurIPS*, 2017.

[51] S. Verdoolaege, "isl: An integer set library for the polyhedral model," in *International Congress on Mathematical Software*, 2010.

[52] S. Verdoolaege, R. Seghir, K. Beyls, V. Loechner, and M. Bruynooghe, "Counting integer points in parametric polytopes using barvinok's rational functions," *Algorithmica*, 2007.

[53] B. Viswanath, A. Mislove, M. Cha, and K. P. Gummadi, "On the evolution of user interaction in facebook," in *Proceedings of the 2nd ACM workshop on Online social networks*, 2009.

[54] J. Wang, L. Guo, and J. Cong, "AutoSA: A Polyhedral Compiler for High-Performance Systolic Arrays on FPGA," in *Proceedings of the 2021 ACM/SIGDA international symposium on Field-programmable gate arrays*, 2021.

[55] M. E. Wolf and M. S. Lam, "A Data Locality Optimizing Algorithm," in *Proceedings of the ACM SIGPLAN 1991 Conference on Programming Language Design and Implementation*, 1991.

[56] X. Yang, M. Gao, Q. Liu, J. Setter, J. Pu, A. Nayak, S. Bell, K. Cao, H. Ha, P. Raina, C. Kozyrakis, and M. Horowitz, "Interstellar: Using halide's scheduling language to analyze dnn accelerators," in *Proceedings of the Twenty-Fifth International Conference on Architectural Support for Programming Languages and Operating Systems*, 2020.

[57] S. Zheng, Y. Liang, S. Wang, R. Chen, and K. Sheng, "Flextensor: An automatic schedule exploration and optimization framework for tensor computation on heterogeneous system," in *Proceedings of the Twenty-Fifth International Conference on Architectural Support for Programming Languages and Operating Systems*, 2020.